\PassOptionsToPackage{unicode}{hyperref}
\PassOptionsToPackage{hyphens}{url}
\PassOptionsToPackage{dvipsnames,svgnames,x11names}{xcolor}
\documentclass[
  sn-mathphys-num,
]{sn-jnl}

\usepackage{amsmath,amssymb}
\usepackage{iftex}
\ifPDFTeX
  \usepackage[T1]{fontenc}
  \usepackage[utf8]{inputenc}
  \usepackage{textcomp} 
\else 
  \usepackage{unicode-math}
  \defaultfontfeatures{Scale=MatchLowercase}
  \defaultfontfeatures[\rmfamily]{Ligatures=TeX,Scale=1}
\fi
\usepackage{lmodern}
\ifPDFTeX\else  
\fi
\IfFileExists{upquote.sty}{\usepackage{upquote}}{}
\IfFileExists{microtype.sty}{
  \usepackage[]{microtype}
  \UseMicrotypeSet[protrusion]{basicmath} 
}{}
\makeatletter
\@ifundefined{KOMAClassName}{
  \IfFileExists{parskip.sty}{%
    \usepackage{parskip}
  }{
    \setlength{\parindent}{0pt}
    \setlength{\parskip}{6pt plus 2pt minus 1pt}}
}{
  \KOMAoptions{parskip=half}}
\makeatother
\usepackage{xcolor}
\makeatletter
\ifx\paragraph\undefined\else
  \let\oldparagraph\paragraph
  \renewcommand{\paragraph}{
    \@ifstar
      \xxxParagraphStar
      \xxxParagraphNoStar
  }
  \newcommand{\xxxParagraphStar}[1]{\oldparagraph*{#1}\mbox{}}
  \newcommand{\xxxParagraphNoStar}[1]{\oldparagraph{#1}\mbox{}}
\fi
\ifx\subparagraph\undefined\else
  \let\oldsubparagraph\subparagraph
  \renewcommand{\subparagraph}{
    \@ifstar
      \xxxSubParagraphStar
      \xxxSubParagraphNoStar
  }
  \newcommand{\xxxSubParagraphStar}[1]{\oldsubparagraph*{#1}\mbox{}}
  \newcommand{\xxxSubParagraphNoStar}[1]{\oldsubparagraph{#1}\mbox{}}
\fi
\makeatother

\providecommand{\tightlist}{%
  \setlength{\itemsep}{0pt}\setlength{\parskip}{0pt}}\usepackage{longtable,booktabs,array}
\usepackage{calc} 
\usepackage{etoolbox}
\makeatletter
\patchcmd\longtable{\par}{\if@noskipsec\mbox{}\fi\par}{}{}
\makeatother
\IfFileExists{footnotehyper.sty}{\usepackage{footnotehyper}}{\usepackage{footnote}}
\makesavenoteenv{longtable}
\usepackage{graphicx}
\makeatletter
\newsavebox\pandoc@box
\newcommand*\pandocbounded[1]{
  \sbox\pandoc@box{#1}%
  \Gscale@div\@tempa{\textheight}{\dimexpr\ht\pandoc@box+\dp\pandoc@box\relax}%
  \Gscale@div\@tempb{\linewidth}{\wd\pandoc@box}%
  \ifdim\@tempb\p@<\@tempa\p@\let\@tempa\@tempb\fi
  \ifdim\@tempa\p@<\p@\scalebox{\@tempa}{\usebox\pandoc@box}%
  \else\usebox{\pandoc@box}%
  \fi%
}
\def\fps@figure{htbp}
\makeatother

\usepackage{svg}
\usepackage{graphicx}%
\usepackage{multirow}%
\usepackage{amsmath,amssymb,amsfonts}%
\usepackage{amsthm}%
\usepackage{mathrsfs}%
\usepackage[title]{appendix}%
\usepackage{xcolor}%
\usepackage{textcomp}%
\usepackage{manyfoot}%
\usepackage{booktabs}%
\usepackage{algorithm}%
\usepackage{algorithmicx}%
\usepackage{algpseudocode}%
\usepackage{listings}%

\usepackage{multirow}
\usepackage{centernot}
\usepackage{mathtools}
\makeatletter
\@ifpackageloaded{caption}{}{\usepackage{caption}}
\AtBeginDocument{%
\ifdefined\contentsname
  \renewcommand*\contentsname{Table of contents}
\else
  \newcommand\contentsname{Table of contents}
\fi
\ifdefined\listfigurename
  \renewcommand*\listfigurename{List of Figures}
\else
  \newcommand\listfigurename{List of Figures}
\fi
\ifdefined\listtablename
  \renewcommand*\listtablename{List of Tables}
\else
  \newcommand\listtablename{List of Tables}
\fi
\ifdefined\figurename
  \renewcommand*\figurename{Figure}
\else
  \newcommand\figurename{Figure}
\fi
\ifdefined\tablename
  \renewcommand*\tablename{Table}
\else
  \newcommand\tablename{Table}
\fi
}
\@ifpackageloaded{float}{}{\usepackage{float}}
\floatstyle{ruled}
\@ifundefined{c@chapter}{\newfloat{codelisting}{h}{lop}}{\newfloat{codelisting}{h}{lop}[chapter]}
\floatname{codelisting}{Listing}

\makeatother
\makeatletter
\@ifpackageloaded{caption}{}{\usepackage{caption}}
\@ifpackageloaded{subcaption}{}{\usepackage{subcaption}}
\makeatother

\usepackage{bookmark}

\usepackage{xurl} 
\hypersetup{
  pdftitle={A Probabilistic Model of Bilateral Lymphatic Spread in Head and Neck Cancer},
  colorlinks=true,
  linkcolor={blue},
  filecolor={Maroon},
  citecolor={Blue},
  urlcolor={Blue},
  pdfcreator={LaTeX via pandoc}}

\title[A Probabilistic Model of Bilateral Lymphatic Spread in Head and
Neck Cancer]{A Probabilistic Model of Bilateral Lymphatic Spread in Head
and Neck Cancer}

\author*[1,2]{\fnm{Roman} \sur{Ludwig}}\email{roman.ludwig@usz.ch}\author[1,2]{\fnm{Yoel Perez} \sur{Haas}}\email{yoel.perezhaas@usz.ch}\author[3]{\fnm{Sergi} \sur{Benavente}}\email{sergi.benavente@vallhebron.cat}\author[2]{\fnm{Panagiotis} \sur{Balermpas}}\email{panagiotis.balermpas@usz.ch}\author[1,2]{\fnm{Jan} \sur{Unkelbach}}\email{jan.unkelbach@usz.ch}
\affil[1]{\orgdiv{Dep. of Physics}, \orgname{University of Zurich}, Switzerland}
\affil[2]{\orgdiv{Dep. of Radiation Oncology}, \orgname{University Hospital
Zurich}, Switzerland}
\affil[3]{\orgdiv{Dep. of Radiation Oncology}, \orgname{University Hospital Vall
d'Hebron}, Spain}

\abstract{Purpose: Current guidelines for elective nodal irradiation in
oropharyngeal squamous cell carcinoma (OPSCC) recommend including large
portions of the contralateral lymphatic system in the clinical target
volume (CTV-N), even for lateralized tumors with no clinical lymph node
involvement in the contralateral neck. This study introduces a
probabilistic model of bilateral lymphatic tumor progression in OPSCC to
estimate personalized risks of occult disease in specific lymph node
levels (LNLs) based on clinical lymph node involvement, T-stage, and
tumor lateralization.

Methods: Building on a previously developed hidden Markov model for
ipsilateral lymphatic spread, we extend the approach to contralateral
neck involvement. The model represents LNLs I, II, III, IV, V, and VII
on both sides of the neck as binary hidden variables (healthy or
involved), connected via arcs representing spread probabilities. These
probabilities are learned using Markov chain Monte Carlo (MCMC) sampling
from a dataset of 833 OPSCC patients, enabling the model to reflect the
underlying lymphatic progression dynamics.

Results: The model accurately and precisely describes observed patterns
of lymph node involvement with a compact set of interpretable
parameters. Midline extension of the primary tumor is identified as the
primary risk factor for contralateral involvement, with advanced T-stage
and extensive ipsilateral involvement further increasing risk. Occult
disease in contralateral LNL III is highly unlikely if upstream LNL II
is clinically negative, and in contralateral LNL IV, occult disease is
exceedingly rare without LNL III involvement.

Conclusions: This model offers an interpretable, probabilistic framework
to inform personalized elective CTV-N volume reduction. For lateralized
tumors that do not cross the midline, it suggests the contralateral neck
may safely be excluded from elective irradiation. For tumors extending
across the midline but with a clinically negative contralateral neck,
elective irradiation could be limited to LNL II, reducing unnecessary
exposure of normal tissue while maintaining regional tumor control.}

\begin{document}
\maketitle

\section{Introduction}\label{introduction}

In head and neck squamous cell carcinomas (HNSCC) treatments with
radiotherapy or surgery, both the primary tumor and clinically detected
lymph node metastases are targeted. In addition, current guidelines
include large portions of the neck in the elective clinical target
volume (CTV-N)
\citep{gregoire_ctbased_2003, gregoire_delineation_2014, gregoire_delineation_2018, eisbruch_intensitymodulated_2002, biau_selection_2019, chao_determination_2002, vorwerk_guidelines_2011, ferlito_elective_2009}
to mitigate the risk of regional recurrences from untreated microscopic
disease undetectable by in-vivo imaging modalities such as computed
tomography (CT), magnetic resonance imaging (MRI), or positron emission
tomography (PET). However, this approach must balance minimizing the
risk of occult disease in the lymphatic drainage region against the
toxicity of unnecessarily irradiating healthy tissue.

These CTV-N guidelines rely on anatomically defined lymph node levels
(LNLs) \citep{gregoire_delineation_2014} and the overall prevalence of
lymph node metastases within these levels. They often recommend
extensive irradiation of both sides of the neck. However, the general
prevalence of metastasis in a given LNL does not correspond to an
individual patient's risk of occult disease in that region, which
depends on their specific state of tumor progression. For example, a
patient with no clinically detectable nodal disease (cN0) who has a
small, clearly lateralized T1 tumor would receive the same contralateral
CTV-N as a patient with significant ipsilateral nodal involvement and an
advanced tumor crossing the mid-sagittal plane. Both patients receive
elective irradiation of the contralateral LNLs II, III, and IVa
\citep{biau_selection_2019}.

To better quantify individualized risk of occult disease, we previously
developed an intuitive probabilistic hidden Markov model (HMM)
\citep{ludwig_hidden_2021, ludwig_modelling_2023}, originally based on a
conceptually similar a Bayesian network model
\citep{pouymayou_bayesian_2019}. However, these models have been limited
to predicting ipsilateral nodal involvement. This work extends the model
to include contralateral risk predictions, enabling more personalized
radiation volume recommendations for the contralateral neck. By
identifying patients with low contralateral risk, the model could guide
reductions in the contralateral CTV-N, thereby decreasing
radiation-induced toxicity and improving quality of life.

The main contributions of this paper are as follows:

\begin{enumerate}
\def\labelenumi{\arabic{enumi}.}
\item
  Section~\ref{sec-data} presents a multi-centric dataset on lymph node
  involvement in 833 OPSCC patients, identifying key risk factors for
  contralateral lymph node involvement and outlining requirements for a
  bilateral model extension (section~\ref{sec-requirements}).
\item
  Section~\ref{sec-ext-to-contra} introduces a bilateral HMM that
  incorporates primary tumor lateralization, T-category, and clinical
  involvement as risk factors for contralateral involvement. Model
  training and computational experiments are described in
  section~\ref{sec-methods}.
\item
  Section~\ref{sec-results} demonstrates the model's ability to
  replicate observed contralateral lymph node involvement patterns and
  estimates occult disease risk for typical patients. Implications for
  volume-deescalated radiotherapy are discussed in
  section~\ref{sec-discussion}.
\end{enumerate}

\section{Data on Lymphatic Progression Patterns}\label{sec-data}

{
\setlength{\tabcolsep}{6pt}
\begin{longtable}[]{@{}lrrrrrr@{}}

\caption{\label{tbl-data-overview}Overview over the four datasets from
four different institutions used to train and evaluate our model. Here,
we briefly characterize the total number of OPSCC patients from the
respective institution, their median age, what proportion received neck
dissection, the N0 portion of patients, what percentage presented with
early T-category (T1/T2), and the prevalence of primary tumor midline
extension. For a much more detailed look at the data, visit
\href{https://lyprox.org}{lyprox.org}.}

\tabularnewline

\tabularnewline
\toprule\noalign{}
Institution & Total & Median Age & Surgery & N0 & T1 or T2
& Mid. Ext. \\
\midrule\noalign{}
\endfirsthead
\toprule\noalign{}
Institution & Total & Median Age & Surgery & N0 & T1 or T2
& Mid. Ext. \\
\midrule\noalign{}
\endhead
\bottomrule\noalign{}
\endlastfoot
CLB (Lyon) & 325 & 60 & 100\% & 19\% & 69\% & 18\% \\
ISB (Bern) & 74 & 61 & 100\% & 18\% & 66\% & 14\% \\
USZ (Zurich) & 287 & 66 & 26\% & 18\% & 52\% & 31\% \\
HVH (Barcelona) & 147 & 58 & 5\% & 21\%
& 34\% & 34\% \\

\end{longtable}
}

To develop models for lymphatic tumor progression for all relevant LNLs,
including contralateral regions, we compiled a detailed dataset of 833
patients with newly diagnosed oropharyngeal squamous cell carcinomas
(OPSCC) \citep{ludwig_dataset_2022, ludwig_multicentric_2024}. The
dataset includes lymph node involvement per LNL for each patient in
tabular form, along with primary tumor and patient characteristics such
as T-category, subsite, primary tumor lateralization, and HPV p16
status. Patient records were collected from four institutions, and an
overview of patient characteristics is provided in
table~\ref{tbl-data-overview}.

Data from the Inselspital Bern (ISB) and Centre Léon Bérard (CLB) consist
exclusively of patients who underwent neck dissections. In contrast, the
majority of patients from the University Hospital Zürich (USZ) and the
Hospital Vall d'Hebron (HVH) were treated with definitive radiotherapy.
Since surgical treatment is more common for early T-category patients,
ISB and CLB datasets include a higher proportion of these cases compared
to USZ and HVH. For 83 patients in the CLB dataset, the primary tumor's
lateralization was not reported.

\subsection{Consensus on Involvement Status}\label{sec-data-consensus}

Pathological involvement is available only for surgically treated
patients and for the levels that were dissected. For non-surgical
patients, involvement status is determined clinically, i.e.~using
imaging. For this work, diagnostic information was synthesized into a
consensus decision for each patient and LNL. This consensus reflects the
most likely state of involvement and accounts for the sensitivity and
specificity of various diagnostic modalities, as reported in the
literature
\citep{debondt_detection_2007, kyzas_18ffluorodeoxyglucose_2008}.

The consensus process is detailed in section~\ref{sec-consensus}.
Briefly, pathological findings from neck dissections are treated as the
gold standard, overriding any conflicting clinical diagnoses. For levels
not dissected, PET-CT is typically the primary source for determining
the most likely state of involvement.

\subsection{Data Availability}\label{data-availability}

This publication pools several datasets on lymphatic progression patterns. Most of them can be downloaded and explored in our interactive web app LyProX (\url{https://lyprox.org}). They are also stored in the lyDATA GitHub repository (\url{https://github.com/rmnldwg/lydata}). Further, they are described in dedicated Data-in-Brief publications \citep{ludwig_dataset_2022,ludwig_multicentric_2024} and deposited in the zenodo archive (\url{https://doi.org/10.5281/zenodo.10204085}, \url{https://doi.org/10.5281/zenodo.10210423}, \url{https://doi.org/10.5281/zenodo.10210361}, and \url{https://doi.org/10.5281/zenodo.5833835}). The remaining patient records will be published at a later time and may be provided upon request.

\subsection{Patterns of Contralateral Involvement}\label{sec-data-strat}

\begin{figure}

\centering{

\pandocbounded{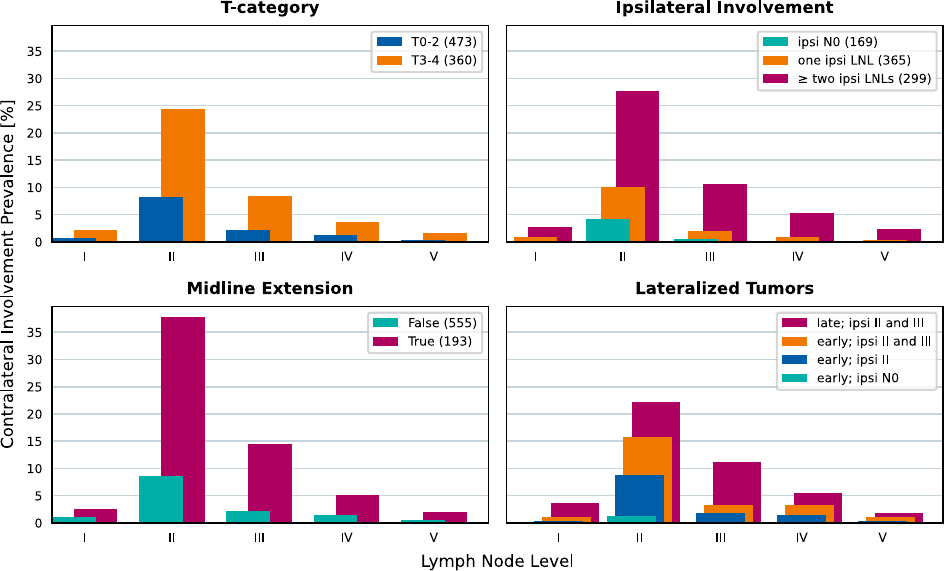}

}

\caption{\label{fig-data-strat}Contralateral involvement stratified by
T-category (top left panel), the number of metastatic LNLs ipsilaterally
(top right panel), and whether the primary tumor extended over the
mid-sagittal plane or was clearly lateralized (bottom left panel). In
the bottom right panel, we consider lateralized tumors only, and compare
the contralateral involvement prevalence for selected scenarios that
vary in their T-category and ipsilateral involvement extent.}

\end{figure}

The datasets enable analysis of correlations between contralateral LNL
involvement and key risk factors. In figure~\ref{fig-data-strat}, we
illustrate the prevalence of contralateral LNL involvement, stratified
by T-category, the number of ipsilaterally involved LNLs, and whether
the tumor extends across the mid-sagittal plane.

\subsubsection{Midline Extension}\label{midline-extension}

The bottom left panel of figure~\ref{fig-data-strat} shows that tumors
crossing the mid-sagittal plane have a substantially higher prevalence
of contralateral involvement compared to clearly lateralized tumors.
This aligns with the anatomy of the head and neck lymphatic system,
which is symmetric, with no major lymph vessels crossing the midline.
Interstitial fluids from the primary tumor, presumed to carry malignant
cells, can merely diffuse to contralateral lymphatic vessels over short
distances. Thus, contralateral spread is more likely when the tumor
approaches or crosses the midline.

\subsubsection{T-Category}\label{t-category}

The top left panel shows a correlation between T-category and
contralateral involvement, reflecting T-category's role as a surrogate
for the time elapsed between disease onset and diagnosis. Advanced
T-category tumors (e.g., T4) generally represent a longer disease
progression timeline, providing more opportunity for metastatic spread
compared to smaller tumors (e.g., T1).

\subsubsection{Ipsilateral Involvement}\label{ipsilateral-involvement}

The top right panel reveals a positive correlation between ipsilateral
and contralateral metastases. Extensive ipsilateral involvement likely
indicates a longer or faster disease progression. Additionally, it has
been hypothesized that bulky ipsilateral nodal disease may reroute
lymphatic drainage toward the contralateral side, potentially increasing
the probability of contralateral metastasis.

\subsubsection{Correlation of Risk
Factors}\label{correlation-of-risk-factors}

Midline extension, T-category, and ipsilateral involvement are
interrelated risk factors for contralateral metastasis. For instance,
45.6\% of advanced T-category tumors exhibit midline extension compared
to 6.1\% of early T-category tumors. While the higher fraction of
midline extensions in advanced T-category patients partially explains
the higher contralateral metastasis rates, T-category itself and
ipsilateral involvement also play an additional role.

The bottom right panel of figure~\ref{fig-data-strat} considers only
patients with lateralized tumors that do not cross the midline. Among
early T-category patients with no ipsilateral nodal involvement (levels
I-V), only 1.2\% (1 of 86 patients) show involvement in contralateral
level II. This proportion increases to 8.8\% (24 of 272) if ipsilateral
level II is involved, to 15.7\% (14 of 89) if ipsilateral levels II and
III are involved, and further to 22.2\% (12 of 54) for advanced
T-category tumors with ipsilateral levels II and III involved.

\subsection{Requirements for a Bilateral Model}\label{sec-requirements}

Based on the observations in section~\ref{sec-data-strat} above, any
model predicting the risk of contralateral nodal involvement should
account for the following:

\begin{enumerate}
\def\labelenumi{\arabic{enumi}.}
\tightlist
\item
  \textbf{Midline Extension:} Tumors extending across the mid-sagittal
  plane should result in a significantly higher probability of
  contralateral metastases.
\item
  \textbf{T-Category:} Advanced T-category should correspond to an
  increased risk of nodal disease. In the hidden Markov model this can
  be modeled using the expected time of diagnosis, as demonstrated
  previously for the ipsilateral model \citep{ludwig_hidden_2021}.
\item
  \textbf{Ipsilateral Involvement:} The model should be able to capture
  the correlation between the extent of ipsi- and contralateral
  involvement. I.e., a more severe ipsilateral involvement should
  indicate a higher risk for contralateral metastases.
\end{enumerate}

\section{Unilateral Model for Lymphatic
Progression}\label{sec-unilateral}

This paper builds on the previously developed unilateral model for
ipsilateral lymph node involvement presented in
\citep{ludwig_modelling_2024}. Below, we briefly recap the unilateral
model to introduce the notation required for extending the framework to
a bilateral model, described in section~\ref{sec-ext-to-contra}. For
further details on the ipsilateral model, we refer to earlier
publications \citep{ludwig_hidden_2021, ludwig_modelling_2024}.

We represent a patient's state of involvement at an abstract time-step
\(t\) as a vector of hidden binary random variables, where each
component corresponds to a lymph node level (LNL):
\begin{equation}\phantomsection\label{eq-state-def}{
\mathbf{X}[t] = \begin{pmatrix} X_v[t] \end{pmatrix} \qquad v \in \left\{ 1, 2, \ldots, V \right\}
}\end{equation}

Here, \(V\) is the number of LNLs the model considers. The values a LNL
may take on are \(X_v[t] = 0\) (\texttt{False}), meaning the LNL \(v\)
is free of metastatic disease, or \(X_v[t] = 1\) (\texttt{True}),
corresponding to the presence of clinically detected metastases (i.e.,
occult or macroscopic disease). In total, there are \(2^V\) distinct
possible lymphatic involvement patterns, which we enumerate from
\(\boldsymbol{\xi}_0 = \begin{pmatrix} 0 & 0 & \cdots & 0 \end{pmatrix}\)
to
\(\boldsymbol{\xi}_{2^V} = \begin{pmatrix} 1 & 1 & \cdots & 1 \end{pmatrix}\).
Each LNL's state is observed via another binary random variable \(Z_v\)
that describes the clinical involvement of a LNL based on imaging:
\(Z_v = 0\) (\texttt{False}) indicates that the LNL \(v\) is healthy
based on clinical diagnosis, and \(Z_v = 1\) (\texttt{True}) indicates
that LNL \(v\) was classified as involved. \(X_v\) and \(Z_v\) are
connected through the sensitivity and specificity of the diagnositc
modality.

Based on this, our HMM is fully described by defining the following
three quantities:

\begin{enumerate}
\def\labelenumi{\arabic{enumi}.}
\tightlist
\item
  A starting state \(\mathbf{X}[t=0]\) at time \(t=0\) just before the
  patient's tumor formed. In our case, this is always the state
  \(\boldsymbol{\xi}_0\) where all LNLs are still healthy.
\item
  The \emph{transition matrix}
  \begin{equation}\phantomsection\label{eq-trans-matrix}{
  \mathbf{A} = \left( A_{ij} \right) = \big( P \left( \mathbf{X}[t+1] = \boldsymbol{\xi}_j \mid \mathbf{X}[t] = \boldsymbol{\xi}_i \right) \big)
  }\end{equation} where the value at row \(i\) and column \(j\)
  represents the probability to transition from state
  \(\boldsymbol{\xi}_i\) to \(\boldsymbol{\xi}_j\) during the time-step
  from \(t\) to \(t+1\). Note that we prohibit self-healing, meaning
  that during a transition, no LNL may change their state from
  \(X_v[t]=1\) to \(X_v[t+1]=0\). Consequently, many elements of the
  transition matrix are zero.
\item
  Lastly, the \emph{observation matrix}
  \begin{equation}\phantomsection\label{eq-obs-matrix}{
  \mathbf{B} = \left( B_{ij} \right) = \big( P \left( \mathbf{Z} = \boldsymbol{\zeta}_j \mid \mathbf{X}[t_D] = \boldsymbol{\xi}_i \right) \big)
  }\end{equation} where in row \(i\) and at column \(j\) we find the
  probability to \emph{observe} a lymphatic involvement pattern
  \(\mathbf{Z} = \boldsymbol{\zeta}_j\), given that the true (but
  hidden) state of involvement at the time of diagnosis \(t_D\) is
  \(\mathbf{X}[t_D] = \boldsymbol{\xi}_i\).
\end{enumerate}

\begin{figure}

\centering{

\def\svgwidth{0.4\linewidth}
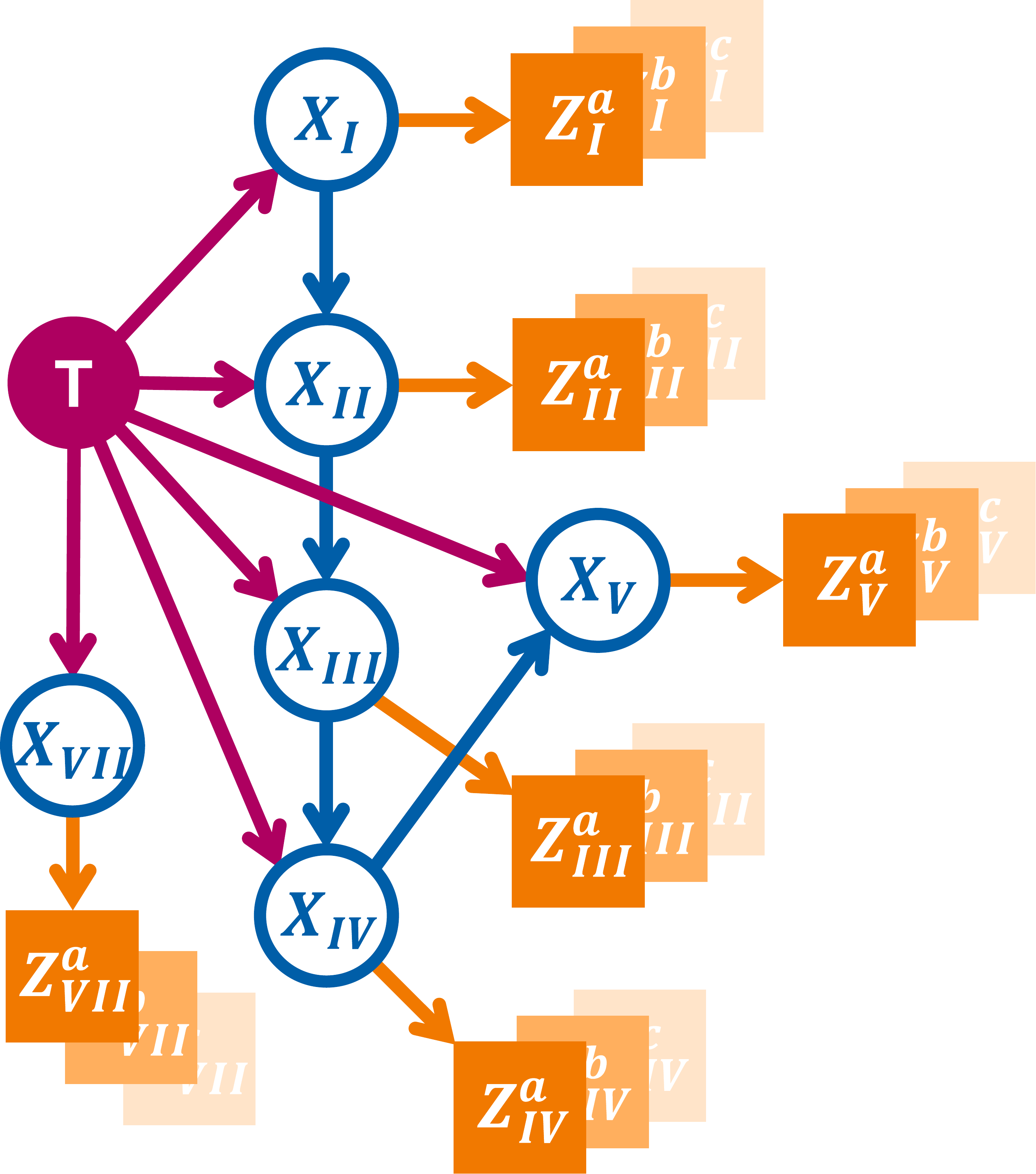

}

\caption{\label{fig-full-graph}Directed acyclic graph (DAG) representing
the abstract lymphatic network in the head and neck region. Blue nodes
are the LNLs' hidden random variables, the red node represents the
tumor, and the orange square nodes depict the binary observed variables.
Red and blue arcs symbolize the probability of lymphatic spread along
that edge during one time-step. The orange arcs represent the
sensitivity and specificity of the observational modality (e.g.~CT, MRI,
pathology, \ldots).}

\end{figure}

The transition matrix \(\mathbf{A}\) is parameterized using a directed
acyclic graph (DAG) that represents the underlying lymphatic network.
Edges from the primary tumor to an LNL are associated with a probability
\(b_v\) for direct spread to LNL \(v\) during one time step. Arcs from a
LNL \(v\) to a LNL \(r\) are parameterized with the probability rate
\(t_{vr}\) representing the probability of spread to an LNL \(r\) that
receives efferent lymphatic spread from LNL \(v\). In this paper, we
build on the DAG shown in figure~\ref{fig-full-graph} which was obtained
by maximizing the model evidence as described in
\citep{ludwig_modelling_2024}.

Let us now consider the probability distribution over all possible
hidden states \(\mathbf{X}[t]\) at time \(t\). We can get to this
distribution by evolving the healthy starting state
\(\mathbf{X}[t=0] = \boldsymbol{\xi}_0\) at time \(t=0\) step by step,
by successively multiplying this vector with the transition matrix
\(\mathbf{A}\): \(\mathbf{X}[t+1] = \mathbf{X}[t] \cdot \mathbf{A}\).
For later use, we define at this point a matrix \(\boldsymbol{\Lambda}\)
that collects these distributions for all considered times:
\begin{equation}\phantomsection\label{eq-lambda-matrix}{
\boldsymbol{\Lambda} = P \left( \mathbf{X} \mid \mathbf{t} \right) = \begin{pmatrix}
\boldsymbol{\pi}^\intercal \cdot \mathbf{A}^0 \\
\boldsymbol{\pi}^\intercal \cdot \mathbf{A}^1 \\
\vdots \\
\boldsymbol{\pi}^\intercal \cdot \mathbf{A}^{t_\text{max}} \\
\end{pmatrix}
}\end{equation}
where the \(k\)-th row in this matrix corresponds to the probability
distribution over hidden states after \(t=k-1\) time-steps.

At the time of diagnosis, \(0 \leq t_D \leq t_\text{max}\), we multiply
the evolved state distribution with the observation matrix
\(\mathbf{B}\) to obtain the distribution over all possible diagnoses.
However, the exact time of diagnosis, \(t_D\), is unknown; that is, we
do not know the number of time-steps over which the HMM should be
evolved. To address this, we marginalize over all possible diagnosis
times, allowing the diagnosis to occur at any time-step, albeit with
different weights. These weights are defined by a prior distribution
over \(t_D\), which can vary depending on the patient's T-category. For
example, the time-prior for early T-category patients,
\(P(t_D \mid \text{early})\), may put more weight on earlier time-steps,
reflecting -- on average -- earlier detection, compared to the prior for
advanced T-category patients, \(P(t_D \mid \text{advanced})\).

The probability distribution over \(\mathbf{X}\) for the example of an
early T-category patient is given by
\[
P\left( \mathbf{X} \mid \text{T}x = \text{early} \right) = \sum_{t=0}^{t_\text{max}} P \left( \mathbf{X} \mid t \right) \cdot P(t \mid \text{T}x = \text{early})
\]

In this work, we use binomial distributions
\(\mathfrak{B} \left( t_D, p_{\text{T}x} \right)\) as time-priors which
have one free parameter \(p_{\text{T}x}\) for each group of patients we
differentiate based on T-category. Also, we fix \(t_\text{max} = 10\),
which means that the expected number of time-steps from the onset of a
patient's disease to their diagnosis is
\(\mathbb{E}\left[ t_D \right] = 10 \cdot p_{\text{T}x}\).

\subsection{Likelihood Function of the Unilateral
Model}\label{likelihood-function-of-the-unilateral-model}

The probabiltiy for a patient to present with a diagnosis
\(\mathbf{Z} = \boldsymbol{\zeta}_i\) a T-category \(\text{T}x\) tumor
can now be written as:
\begin{equation}\phantomsection\label{eq-single-patient-llh}{
\ell = P \left( \mathbf{Z} = \boldsymbol{\zeta}_i \mid \text{T}x \right) = \sum_{t=0}^{t_\text{max}} \left[ \boldsymbol{\xi}_0 \cdot \mathbf{A}^t \cdot \mathbf{B} \right]_i \cdot P \left( t \mid \text{T}x \right)
}\end{equation}
here, \(\left[ \ldots \right]_i\) we denote the \(i\)-th component of
the vector in the square brackets. Note that it is also possible to
account for missing involvement information: If a diagnosis (like fine
needle aspiration (FNA)) is only available for a subset of all LNLs, we
can sum over all those possible complete observed states
\(\boldsymbol{\zeta}_j\) that match the provided diagnosis.

The term above represents a single patient's contribution to the overall
likelihood function that is a product of such terms for each patient.
This single-patient likelihood \(\ell\) in
equation~\ref{eq-single-patient-llh} depends on the spread parameters
shown in figure~\ref{fig-full-graph} via the transition matrix
\(\mathbf{A}\) and on the binomial parameters \(p_{\text{T}x}\) via
time-priors. In this work, we will only differentiate between ``early''
(T1 \& T2) and ``advanced'' (T3 \& T4) T-categories. Therefore, the
parameter space of the unilateral model is:
\begin{equation}\phantomsection\label{eq-param-space}{
\boldsymbol{\theta} = \left( \left\{ b_v \right\}, \left\{ t_{vr} \right\}, p_\text{early}, p_\text{adv.} \right) \quad \text{with} \quad \genfrac{}{}{0pt}{2}{v\leq V}{r\in\operatorname{pa}(v)}
}\end{equation}
And it is our goal to infer optimal parameter values of from a given
dataset \(\mathcal{D}\) (consisting of diagnoses and T-categories) of
OPSCC patients. The likelihood to observe this cohort of \(N\) patients,
given a set of parameters \(\boldsymbol{\theta}\) is simply the product
of their individual likelihoods as defined in
equation~\ref{eq-single-patient-llh}. For numerical reasons, we
typically compute the data likelihood in log space:
\begin{equation}\phantomsection\label{eq-log-likelihood}{
\log \mathcal{L} \left( \mathcal{D} \mid \boldsymbol{\theta} \right) = \sum_{i=1}^N \log \ell_i
}\end{equation}
The methodology we use to infer the model's parameters is detailed in
section~\ref{sec-sampling}.

\section{Extension to a Bilateral Model}\label{sec-ext-to-contra}

A straightforward approach to modeling contralateral lymphatic spread
would be to use two independent unilateral models, as described in
section~\ref{sec-unilateral}, possibly with shared parameters such as
the distribution of diagnosis times or spread between LNLs (\(t_{vr}\)).
However, this method would fail to capture the correlation between
ipsilateral and contralateral involvement discussed in
section~\ref{sec-data-strat}, particularly the observed increase in
contralateral involvement with greater severity of ipsilateral spread.

Thus, we extend the formalism in section~\ref{sec-unilateral} in such a
way that the model's ipsi- and contralateral side evolve synchronously
over time. To achieve that, we start by writing down the posterior
distribution of involvement, which is now a joint probability of an
involvement \(\mathbf{X}^\text{i}\) ipsilaterally \emph{and} an
involvement \(\mathbf{X}^\text{c}\) contralaterally, given a diagnosis
of the ipsilateral LNLs \(\mathbf{Z}^\text{i}\) and of the contralateral
ones \(\mathbf{Z}^\text{c}\):
\begin{equation}\phantomsection\label{eq-bilateral-bayes}{
P \left( \mathbf{X}^\text{i}, \mathbf{X}^\text{c} \mid \mathbf{Z}^\text{i}, \mathbf{Z}^\text{c} \right) = \frac{P \left( \mathbf{Z}^\text{i}, \mathbf{Z}^\text{c} \mid \mathbf{X}^\text{i}, \mathbf{X}^\text{c} \right) P \left( \mathbf{X}^\text{i}, \mathbf{X}^\text{c} \right)}{P \left( \mathbf{Z}^\text{i}, \mathbf{Z}^\text{c} \right)}
}\end{equation}
For the sake of brevity, we omit the dependency on the parameters and
the T-category here.

The probability of the diagnoses given a hidden state factorises:
\(P \left( \mathbf{Z}^\text{i}, \mathbf{Z}^\text{c} \mid \mathbf{X}^\text{i}, \mathbf{X}^\text{c} \right) = P \left( \mathbf{Z}^\text{i} \mid \mathbf{X}^\text{i} \right) \cdot P \left( \mathbf{Z}^\text{c} \mid \mathbf{X}^\text{c} \right)\),
and the two factors are described through observation matrices
\(\mathbf{B}^\text{i}\) and \(\mathbf{B}^\text{c}\).

The term representing the model's prior probability of hidden
involvement does not factorize. However, we assume no direct lymphatic
drainage from ipsilateral to contralateral LNLs, as major lymph vessels
do not cross the mid-sagittal plane. In the graphical model, this
translates to the absence of directed arcs between ipsilateral and
contralateral LNLs, implying that contralateral tumor spread occurs
solely via the primary tumor. We can thus write the joint probability
\(P \left( \mathbf{X}^\text{i}, \mathbf{X}^\text{c} \right)\) as a
factorising sum:
\begin{equation}\phantomsection\label{eq-bilateral-marginal}{
\begin{aligned}
P \left( \mathbf{X}^\text{i}, \mathbf{X}^\text{c} \right) &= \sum_{t=0}^{t_\text{max}} P(t) \cdot P \left( \mathbf{X}^\text{i}, \mathbf{X}^\text{c} \mid t \right) \\
&= \sum_{t=0}^{t_\text{max}} P(t) \cdot P \left( \mathbf{X}^\text{i} \mid t \right) \cdot P \left( \mathbf{X}^\text{c} \mid t \right)
\end{aligned}
}\end{equation}
This assumption is intuitive: since no major lymph vessels cross the
midline, the ipsilateral and contralateral sides of the lymphatic
network evolve independently over time. However, they are indirectly
coupled through time. For example, a joint state with severe
contralateral involvement and limited ipsilateral involvement is
improbable: Severe contralateral involvement typically occurs at later
time steps, when limited ipsilateral involvement is unlikely.

Using equation~\ref{eq-bilateral-marginal} along with
equation~\ref{eq-lambda-matrix}, we can write the above distribution
algebraically as a product:
\begin{equation}\phantomsection\label{eq-bilateral-marginal-algebra}{
P \left( \mathbf{X}^\text{i} = \boldsymbol{\xi}_n, \mathbf{X}^\text{c} = \boldsymbol{\xi}_m \right) = \left[ \boldsymbol{\Lambda}^\intercal_\text{i} \cdot \operatorname{diag} P(\mathbf{t}) \cdot \boldsymbol{\Lambda}_\text{c} \right]_{n,m}
}\end{equation}

\subsection{Parameter Symmetries}\label{sec-params-symmetry}

The matrices \(\boldsymbol{\Lambda}_\text{i}\) and
\(\boldsymbol{\Lambda}_\text{c}\) could, in principle, be parameterized
with entirely separate parameters, allowing ipsilateral and
contralateral spread rates to differ substantially. However, we simplify
the parameter space by sharing parameters between the two sides, based
on the following three assumptions:

\begin{enumerate}
\def\labelenumi{\arabic{enumi}.}
\tightlist
\item
  \textbf{Shared Graph Structure}: Both ipsilateral and contralateral
  spread are described by the same graph shown in
  figure~\ref{fig-full-graph}.
\item
  \textbf{Symmetric Spread Among LNLs}: The spread among LNLs is assumed
  to be the same on both sides, reflecting the symmetric structure of
  the lymphatic system. Consequently, the spread rates between nodes
  should also be symmetric. This is formalized as:\\
  \begin{equation}\phantomsection\label{eq-symmetries}{
    t_{rv}^\text{c} = t_{rv}^\text{i}
    }\end{equation} for all \(v \leq V\) and
  \(r \in \operatorname{pa}(v)\) being all nodes that spread to \(v\).
\item
  \textbf{Asymmetric Spread from Tumor}: The probabilities of direct
  spread from the primary tumor are clearly different for the ipsi- and
  contralateral neck. In addition, tumor spread to the contralateral
  side varies depending on whether the tumor crosses the mid-sagittal
  plane. This would result in three sets of rates for tumor spread to
  the LNLs: (1) the spread to ipsilateral LNLs \(b^\text{i}_v\), (2) the
  spread to contralateral LNLs as long as the tumor is lateralized
  \(b_v^{\text{c},\epsilon=\texttt{False}}\), (3) the spread to
  contralateral LNLs when the tumor crosses the midline
  \(b_v^{\text{c},\epsilon=\texttt{True}}\). In this work, however, we
  chose to define the latter set as a linear mix of ipsilateral tumor
  spread and contralateral spread in case of a clearly lateralized
  tumor. Thus, using \(\alpha \in [0,1]\) as this mixing parameter, we
  have \begin{equation}\phantomsection\label{eq-mixing}{
    b_v^{\text{c},\epsilon=\texttt{True}} = \alpha \cdot b_v^\text{i} + (1 - \alpha) \cdot b_v^{\text{c},\epsilon=\texttt{False}}
    }\end{equation} In this way, the tumor's midline extension causes
  the contralateral spread to become more like the spread to the
  ipsilateral side.
\end{enumerate}

The full parameter space of this model is now:
\begin{equation}\phantomsection\label{eq-bi-param-space}{
\boldsymbol{\theta} = \left( \left\{ b_v^\text{i} \right\}, \left\{ b_v^\text{c} \right\}, \alpha, \left\{ t_{vr} \right\}, p_\text{early}, p_\text{adv.} \right) \quad \text{with} \quad \genfrac{}{}{0pt}{2}{v\leq V}{r\in\operatorname{pa}(v)}
}\end{equation}
This results in less than a doubling of parameters compared to the
unilateral model. From these parameters, we construct three transition
matrices: the unchanged \(\mathbf{A}_\text{i}\) for the ipsilateral
side, \(\mathbf{A}_\text{c}^{\epsilon=\texttt{False}}\) for
contralateral progression while the tumor is lateralized, and
\(\mathbf{A}_\text{c}^{\epsilon=\texttt{True}}\) for cases where the
tumor crosses the mid-sagittal plane.

\subsection{Modelling Midline Extension}\label{sec-midline}

Most tumors crossing the midline at the time of diagnosis likely began
as lateralized tumors that grew over the midline at a later point in
time. As a result, the transition matrix
\(\mathbf{A}_\text{c}^{\epsilon=\texttt{True}}\) applies only to a
subset of time-steps.

To account for this, we model the tumor's extension over the
mid-sagittal plane as an additional binary random variable \(\epsilon\).
A tumor starts as lateralized, with a finite probability \(p_\epsilon\)
at each time step of crossing the midline. The overall probabilities of
a patient having a clearly lateralized tumor or one extending over the
mid-sagittal plane after \(t\) time steps are then given by
\[
\begin{aligned}
P(\epsilon = \texttt{False} \mid t) &= (1 - p_\epsilon)^t \\
P(\epsilon = \texttt{True} \mid t) &= 1 - P(\epsilon = \texttt{False} \mid t)
\end{aligned}
\]

Using this, it is straightforward to write down the matrix of state
distributions for all time-steps, as in equation~\ref{eq-lambda-matrix},
covering the joint distribution over the contralateral hidden state and
the midline extension:

\[
\boldsymbol{\Lambda}_\text{c}^{\epsilon=\texttt{False}} =
\left(
\begin{array}{r}
\boldsymbol{\pi}^\intercal \cdot \left( \mathbf{A}_\text{c}^{\epsilon=\texttt{False}} \right)^{0\phantom{t_\text{max}}} \\
(1-p_\epsilon) \cdot \boldsymbol{\pi}^\intercal \cdot \left( \mathbf{A}_\text{c}^{\epsilon=\texttt{False}} \right)^{1\phantom{t_\text{max}}} \\
\hfill \vdots \hfill \\
(1-p_\epsilon)^{t_\text{max}} \cdot \boldsymbol{\pi}^\intercal \cdot \left( \mathbf{A}_\text{c}^{\epsilon=\texttt{False}} \right)^{t_\text{max}\phantom{0}} \\
\end{array}
\right)
\]

here, we used the transition matrix
\(\mathbf{A}_\text{c}^{\epsilon=\texttt{False}}\) that depends on the
base spread parameters \(b_v^{\text{c},\epsilon=\texttt{False}}\).

The case of midline extension is more complex: we already marginalize
over the exact time step when the tumor grows over the mid-sagittal
plane. However, at the point of crossing, the contralateral transition
matrix must switch to the increased spread rates,
\(b_v^{\text{c}, \epsilon=\texttt{True}}\), as defined by the linear
mixing in equation~\ref{eq-mixing}. To correctly perform this
marginalization, we iteratively construct the joint distribution
\(P \left( \mathbf{X}^\text{c}, \epsilon=\texttt{True} \mid t \right)\).

We begin at \(t=0\), where all contralateral LNLs are healthy (i.e.,
\(\mathbf{X}_\text{c}=\boldsymbol{\xi}_0\)) and the tumor is lateralized
(\(\epsilon=\texttt{False}\)):

\[
P \left( \mathbf{X}^\text{c} = \boldsymbol{\xi}_0, \epsilon=\texttt{False} \mid t=0 \right) = 1
\]

while all other states have zero probability.

At some later time step \(t=\tau+1\), there are two scenarios to
marginalize over:

\begin{enumerate}
\def\labelenumi{\arabic{enumi}.}
\item
  \textbf{The tumor was lateralized at \(t=\tau\) and grew over the
  midline at \(t=\tau+1\):}\\
  In this case, the probability of midline extension at \(t=\tau+1\) is
  \(p_\epsilon\). This probability weights the contralateral state
  distribution that had previously evolved without increased
  contralateral spread.
\item
  \textbf{The tumor had already crossed the midline before
  \(t=\tau\):}\\
  Here, the tumor remains in the midline-crossed state with probability
  1. To account for this scenario, we simply include the distribution
  \(P\left( \mathbf{X}^\text{c}, \epsilon=\texttt{True} \mid \tau \right)\)
  from the previous time step.
\end{enumerate}

Combining these scenarios leads to a recursive formulation:

\[
\begin{multlined}
P \left( \mathbf{X}^\text{c}, \epsilon=\texttt{True} \mid \tau + 1 \right) \\
= \big[ p_\epsilon P \left( \mathbf{X}^\text{c}, \epsilon=\texttt{False} \mid \tau \right) + P \left( \mathbf{X}^\text{c}, \epsilon=\texttt{True} \mid \tau \right) \big]^\top \cdot \mathbf{A}_\text{c}^{\epsilon=\texttt{True}}
\end{multlined}
\]

We can collect the iteratively computed distributions for the midline
extension case to define the matrix over the states given all
time-steps, as in equation~\ref{eq-lambda-matrix}:

\[
\boldsymbol{\Lambda}_\text{c}^{\epsilon=\texttt{True}} = \begin{pmatrix}
P \left( \mathbf{X}^\text{c}, \epsilon=\texttt{True} \mid 0 \right) \\
P \left( \mathbf{X}^\text{c}, \epsilon=\texttt{True} \mid 1 \right) \\
\vdots \\
P \left( \mathbf{X}^\text{c}, \epsilon=\texttt{True} \mid t_\text{max} \right) \\
\end{pmatrix}
\]

In analogy to equation~\ref{eq-bilateral-marginal-algebra}, we can now
write the joint distribution of ipsi- and contralateral involvement and
midline extension algebraically:
\begin{equation}\phantomsection\label{eq-midline-marginal-algebra}{
P \left( \mathbf{X}^\text{i} = \boldsymbol{\xi}_n, \mathbf{X}^\text{c} = \boldsymbol{\xi}_m, \epsilon \right) = \left[ \boldsymbol{\Lambda}^\intercal_\text{i} \cdot \operatorname{diag} P(\mathbf{t}) \cdot \boldsymbol{\Lambda}_\text{c}^\epsilon \right]_{n,m}
}\end{equation}
With the above, we compute the likelihood of all patients with and
without midline extension separately. And if for some patients the
information of tumor lateralization is not available, we can simply
marginalize over the unknown variable
\(\epsilon \in \{ \texttt{False}, \texttt{True} \}\).

The final parameter space of our extended model has now reached this
size:
\begin{equation}\phantomsection\label{eq-ext-param-space}{
\boldsymbol{\theta} = \left( \left\{ b_v^\text{i} \right\}, \left\{ b_v^\text{c} \right\}, \alpha, \left\{ t_{vr} \right\}, p_\text{early}, p_\text{adv.}, p_\epsilon \right) \quad \text{with} \quad \genfrac{}{}{0pt}{2}{v\leq V}{r\in\operatorname{pa}(v)}
}\end{equation}

\subsection{Model Prediction in the Bayesian
Context}\label{model-prediction-in-the-bayesian-context}

Our stated goal is to compute the risk for a patient's true ipsi- and
contralateral nodal involvement states \(\mathbf{X}^\text{i}\) and
\(\mathbf{X}^\text{c}\), \emph{given} their individual diagnosis
\(d = \left( \boldsymbol{\zeta}^\text{i}_k, \boldsymbol{\zeta}^\text{c}_\ell, \epsilon, \text{T}x \right)\).
Here, this diagnosis consists of the observed ipsi- and contralateral
nodal involvements, the patient's midline extension \(\epsilon\), and
their tumor's T-category \(\text{T}x\). Using Bayes' law, we can write
this risk as:
\begin{equation}\phantomsection\label{eq-uni-bayes-law}{
P \big( \mathbf{X}^\text{i}, \mathbf{X}^\text{c} \mid d, \boldsymbol{\hat{\theta}} \big)
= \frac{P \left( \boldsymbol{\zeta}^\text{i}_k \mid \mathbf{X}^\text{i} \right) P \left( \boldsymbol{\zeta}^\text{c}_\ell \mid \mathbf{X}^\text{c} \right) P \big( \mathbf{X}^\text{i}, \mathbf{X}^\text{c}, \epsilon \mid \boldsymbol{\hat{\theta}}, \text{T}x \big)}
{\sum_{i=0}^{2^V} \sum_{j=0}^{2^V} \mathcal{C}_{ij}}
}\end{equation}
with the normalization constants

\[
\mathcal{C}_{ij} = P \left( \boldsymbol{\zeta}^\text{i}_k \mid \mathbf{X}^\text{i}=\boldsymbol{\xi}^\text{i}_i \right) P \big( \boldsymbol{\zeta}^\text{c}_\ell \mid \mathbf{X}^\text{c}=\boldsymbol{\xi}^\text{c}_j \big) P \big( \mathbf{X}^\text{i}=\boldsymbol{\xi}^\text{i}_i, \mathbf{X}^\text{c}=\boldsymbol{\xi}^\text{c}_j, \epsilon \mid \boldsymbol{\hat{\theta}}, \text{T}x \big)
\]

The terms
\(P \left( \boldsymbol{\zeta}^\text{i}_k \mid \mathbf{X}^\text{i} \right)\)
and
\(P \left( \boldsymbol{\zeta}^\text{c}_\ell \mid \mathbf{X}^\text{c} \right)\)
are defined solely by sensitivity and specificity of the diagnostic
modality. These terms already appeared in the definition of the
observation matrx in equation~\ref{eq-obs-matrix}. The \emph{prior}
\(P \big( \mathbf{X}^\text{i}, \mathbf{X}^\text{c}, \epsilon \mid \boldsymbol{\hat{\theta}}, \text{T}x \big)\)
in the above equation is the crucial term that is supplied by a trained
model and its parameters \(\boldsymbol{\hat{\theta}}\).

It is possible to compute this \emph{posterior} probability of true
involvement not only for one fully defined state
\((\mathbf{X}^\text{i}, \mathbf{X}^\text{c})\), but also for
e.g.~individual LNLs: For example, the risk for involvement in the
contralateral level IV would be a marginalization over all combination
of ipsi- states \(\boldsymbol{\xi}^\text{i}_i\) contralateral states
\(\boldsymbol{\xi}^\text{c}_j\) where \(\xi^\text{c}_{j4}=1\). Formally:
\begin{equation}\phantomsection\label{eq-marg-over-posterior}{
\begin{multlined}
P \big( \text{IV}^\text{c} \mid \mathbf{Z}^\text{i}=\boldsymbol{\zeta}^\text{i}_k, \mathbf{Z}^\text{c}=\boldsymbol{\zeta}^\text{c}_\ell, \boldsymbol{\hat{\theta}}, \text{T}x \big) \\
= \sum_k \sum_{\ell \, : \, \xi_{\ell 4}=1} P \big( \mathbf{X}^\text{i} = \boldsymbol{\xi}^\text{i}_k, \mathbf{X}^\text{c} = \boldsymbol{\xi}^\text{c}_\ell \mid \boldsymbol{\zeta}^\text{i}_k, \boldsymbol{\zeta}^\text{c}_\ell, \epsilon, \boldsymbol{\hat{\theta}}, \text{T}x \big)
\end{multlined}
}\end{equation}

\section{Computational Methods}\label{sec-methods}

This section details the experimental setup. All figures, tables, and
results are fully reproducible via the GitHub repository
\href{https://github.com/rmnldwg/bilateral-paper}{rmnldwg/bilateral-paper}.

\subsection{Training Data}\label{sec-training-data}

We trained the model using the dataset of 833 patients described in
section~\ref{sec-data}. The consensus decision on lymphatic involvement
is assumed to correspond to the true hidden state of involvement
\(\mathbf{X}\). Patients with T1 and T2 category tumors have been
grouped into an ``early'' T-category group, those with T3 and T4 tumors
into the ``advanced'' T-category group.

\subsection{MCMC Sampling}\label{sec-sampling}

We used the Python package
\href{https://emcee.readthedocs.io/en/stable/}{\texttt{emcee}}
\citep{foreman-mackey_emcee_2013} for parameter inference, implementing
efficient MCMC sampling with parallel affine-invariant samplers. The
sampling algorithms employed differential evolution moves
\citep{terbraak_differential_2008, nelson_run_2013}, with the likelihood
implemented by our
\href{https://lymph-model.readthedocs.io/en/stable/}{\texttt{lymph-model}}
Python package.

We initialized 12 parallel samplers (``walkers'') with random values
from the unit cube, effectively representing a uniform prior
distribution over the model parameters. Convergence was determined by
two criteria:

\begin{enumerate}
\def\labelenumi{\arabic{enumi}.}
\tightlist
\item
  The change in autocorrelation time was less than 5.0e-2.
\item
  The autocorrelation estimate dropped below \(n\) / 50, where \(n\) is
  the chain length. Earlier autocorrelation estimates might not be
  trustworthy.
\end{enumerate}

Samples from this \emph{burn-in phase} before convergence were
discarded. After that, we drew 10 additional samples, spaced 10 steps
apart.

\begin{figure}

\centering{

\pandocbounded{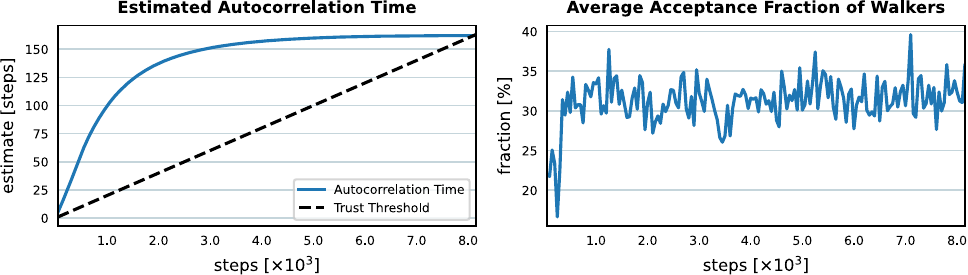}

}

\caption{\label{fig-model-burnin-history}Burn-in phase monitoring of
MCMC sampling. Left: Estimated autocorrelation time, indicating
converging when stable and below the trust threshold. Right: Average
acceptance fraction of parallel walkers, with \textasciitilde30\%
indicating good mixing.}

\end{figure}

We verified sampling convergence in
figure~\ref{fig-model-burnin-history} by examining the MCMC chain's
autocorrelation time and walker acceptance fractions.

\subsection{Comparing the Observed and Predicted Prevalence of
Involvement Patterns}\label{sec-prevalence}

We evaluate the model's ability to describe the observed frequencies of
lymphatic involvement patterns. We compare the prevalence of selected
involvement patterns in the data to the model's predicted prevalence,
given patient scenarios. A ``scenario'' includes the patient's
T-category \(\text{T}x\) and whether the tumor extended over the
mid-sagittal plane, i.e.~\(\epsilon=\texttt{True}\) or
\(\epsilon=\texttt{False}\). An involvement pattern specifies all ipsi-
and contralateral LNLs' status as ``healthy'', ``involved'', or
``masked'' (ignored).

For example, in figure~\ref{fig-model-prevalences-overall} (top left
panel) we assess contralateral LNL II involvement prevalence for early
T-category (T0-T2) and no midline extension
(\(\epsilon=\texttt{False}\)). In the data, \(n=\) 379 such patients
were observed, with \(k=\) 27 exhibiting contralateral LNL II
involvement -- an observed prevalence of \(q=\) 7.1\%. To visualize the
data prevalence, we plot a \emph{beta posterior} over \(q\) -- with a
uniform beta prior -- multiplied with the binomial likelihood for \(k\)
out of \(n\) patients, given \(q\). The resulting distribution has its
maximum at \(q=k / n\) and nicely captures the statistical uncertainty
in the observed cohort: In the top right panel of
figure~\ref{fig-model-prevalences-overall}, we consider the case of
early T-category tumors extending over the midline. The dataset contains
only 29 such patients, out of which 6 had contralateral LNL II
involvement. Consequently, the beta distribution over the observed
prevalence is much wider.

The model's predicted prevalence to compare it with is computed as:
\[
\begin{multlined}
P \left( \text{II}^\text{c} \mid \epsilon=\texttt{False}, \text{T}x=\text{early} \right) \\ = \frac{ \sum_k \sum_{\ell \, : \, \xi_{\ell 2}=1}  P \left( \mathbf{X}^\text{i}=\boldsymbol{\xi}_k^\text{i}, \mathbf{X}^\text{c}=\boldsymbol{\xi}_\ell^\text{c}, \epsilon=\texttt{False} \mid \text{T}x=\text{early} \right)}{ \sum_k \sum_\ell P \left(\mathbf{X}^\text{i}=\boldsymbol{\xi}_k^\text{i}, \mathbf{X}^\text{c}=\boldsymbol{\xi}_\ell^\text{c}, \epsilon=\texttt{False} \mid \text{T}x=\text{early} \right)}
\end{multlined}
\]

In the enumator, we marginalize over all combinations of states in both
sides of the neck where the contralateral LNL II is involved. This is
similar to the marginalization in equation~\ref{eq-marg-over-posterior},
although we are summing over different quantities. In the denominator,
we simply sum out all LNL involvement, leaving only the joint
distribution over midline extension and diagnose time
\(P \left( \epsilon, t \right)\) marginalized over \(t\) using the early
T-category's time-prior.

We display model predictions as histograms, each value computed from one
of the MCMC samples. Ideally, these approximate the location and width
of the beta posteriors from the data showing an accurate and precise
fit.

Note that we omit the y-axis in these figures, as their numerical value
is not intuitively interpretable. We instead use the free space to label
e.g.~rows in an array of subplots.

\section{Results: Model evaluation}\label{sec-results}

In table~\ref{tbl-midline-params}, we tabulate the mean and standard
deviation of the sampled parameters. The bilateral model mostly
reproduces the ipsilateral spread parameter values reported in the
earlier publication on the unilateral model
\citep{ludwig_modelling_2023}. Any discrepancies may be due to
differences in the patient cohorts. Therefore, we omit the analysis of
the ipsilateral spread patterns and focus instead on the analysis of
contralateral involvement. The small contralateral spread parameters
\(b^c_v\) compared to the ipsilateral parameters \(b^i_v\) adequately
reflect the low prevalence of contralateral lymph node involvement for
lateralized tumors. The mixing parameter \(\alpha\) (33.9\%) describes
that the probability of contralateral spread is higher for tumors
extending over the mid-sagittal plane, but still lower than the
probability for ipsilateral spread.

The high value of \(t_{23}\) (14.2\%) compared to the value of
\(b_3^\text{i}\) (5.5\%) shows that the ipsilateral LNL III is rarely
involved without the upstream involvement of LNL II.

\begin{longtable}[]{@{}lrr@{}}

\caption{\label{tbl-midline-params}Mean sampled parameter estimates of
the midline model and the respective standard deviation. The parameters
set to fixed values are the maximum number of time steps \(t_{max}=10\)
and the time prior parameter for early T-category patients
\(p_{early}=0.3\).}

\tabularnewline

\tabularnewline
\toprule\noalign{}
Parameter & Mean & Std. Dev. \\
\midrule\noalign{}
\endfirsthead
\toprule\noalign{}
Parameter & Mean & Std. Dev. \\
\midrule\noalign{}
\endhead
\bottomrule\noalign{}
\endlastfoot
$p_\epsilon$ & 8.16\% & ± 0.48\% \\
$b_1^\text{i}$ & 2.80\% & ± 0.26\% \\
$b_2^\text{i}$ & 34.89\% & ± 1.40\% \\
$b_3^\text{i}$ & 5.45\% & ± 0.66\% \\
$b_4^\text{i}$ & 0.94\% & ± 0.18\% \\
$b_5^\text{i}$ & 1.83\% & ± 0.22\% \\
$b_7^\text{i}$ & 2.32\% & ± 0.26\% \\
$b_1^{\text{c},\epsilon=\texttt{False}}$ & 0.29\% & ± 0.09\% \\
$b_2^{\text{c},\epsilon=\texttt{False}}$ & 2.46\% & ± 0.29\% \\
$b_3^{\text{c},\epsilon=\texttt{False}}$ & 0.14\% & ± 0.07\% \\
$b_4^{\text{c},\epsilon=\texttt{False}}$ & 0.19\% & ± 0.08\% \\
$b_5^{\text{c},\epsilon=\texttt{False}}$ & 0.05\% & ± 0.04\% \\
$b_7^{\text{c},\epsilon=\texttt{False}}$ & 0.50\% & ± 0.17\% \\
$\alpha$ & 33.87\% & ± 4.32\% \\
$t_{12}$ & 62.50\% & ± 16.69\% \\
$t_{23}$ & 14.23\% & ± 1.64\% \\
$t_{34}$ & 15.86\% & ± 1.93\% \\
$t_{45}$ & 14.58\% & ± 3.76\% \\
$p_\text{adv.}$ & 44.98\% & ± 1.99\% \\

\end{longtable}

\subsection{Illustration of the model}\label{illustration-of-the-model}

\begin{figure}

\centering{

\def\svgwidth{0.5\linewidth}
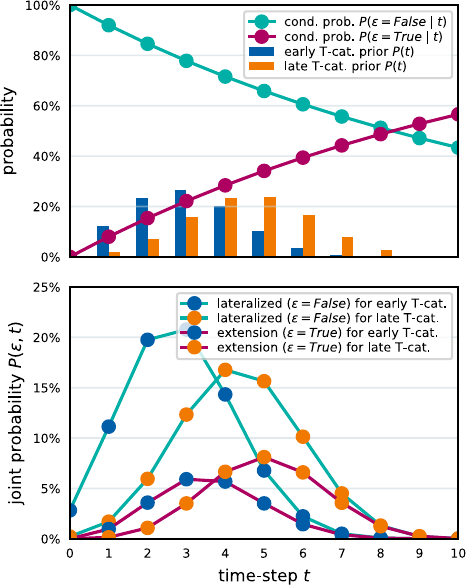

}

\caption{\label{fig-model-midext-evo}The top panel shows the prior
probability to be diagnosed at time step \(t\) for early and late
T-category tumors as bars. The conditional probability of midline
extension (\(\epsilon=\texttt{True}\)) given time step \(t\) is shown as
a line plot. The bottom panel illustrates the joint probability of being
diagnosed at time \(t\) and having a tumor that crosses the midline.}

\end{figure}

In this subsection, we illustrate key aspects of the mathematical
framework introduced earlier. The top panel of
figure~\ref{fig-model-midext-evo} shows the prior distribution over
diagnosis times, \(P(t)\). Based on the parameterization, early
T-category tumors are on average diagnosed after 3 time steps, while
advanced T-category tumors are diagnosed later, averaging 4.5 time
steps. This is due to the learned value of \(p_\text{adv.}\), which is
45\% for advanced T-category tumors. The tumor's average probability per
time step of growing over the midline, \(p_\epsilon\), was found to be
8.2\%. Using this value, the conditional probability of midline
extension, \(P(\epsilon \mid t)\), can be computed for a given time step
\(t\) (red line in the top panel of figure~\ref{fig-model-midext-evo}).
The bottom panel visualizes the joint probability \(P(\epsilon, t)\),
showing the likelihood of diagnosis at time \(t\) with specific states
of midline extension and T-category.

\begin{figure}

\centering{

\pandocbounded{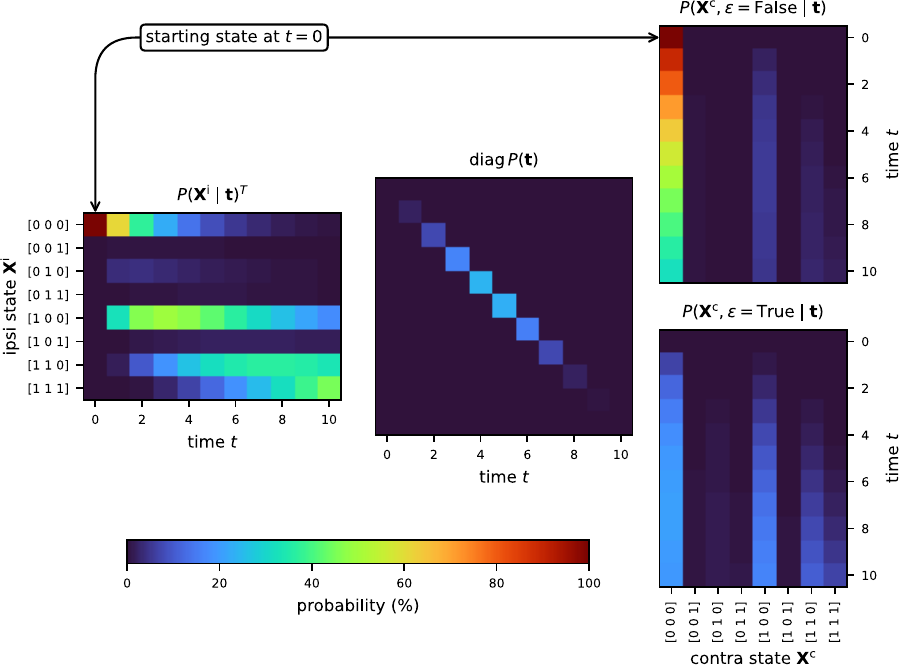}

}

\caption{\label{fig-model-algebra-equation}Visual representation of
equation~\ref{eq-midline-marginal-algebra}. The left and right matrices
represent the time evolution of hidden states for the ipsi- and
contralateral necks, respectively. The right matrices distinguish
between the cases of no midline extension (top) and midline extension
(bottom). The central diagonal matrix shows the time-prior for late
T-category tumors. This computation yields the joint distribution
\(P \left( \mathbf{X}^\text{i}, \mathbf{X}^\text{c}, \epsilon \right)\),
visualized in figure~\ref{fig-model-state-dist}.}

\end{figure}

The framework models the joint probability distribution of midline
extension and ipsi- and contralateral lymph node involvement,
\(P \left( \mathbf{X}^\text{i}, \mathbf{X}^\text{c}, \epsilon \right)\).
This is visualized in figure~\ref{fig-model-algebra-equation}, which
represents the calculation defined in
equation~\ref{eq-midline-marginal-algebra}. To simplify the
interpretation, the example focuses only on LNLs II, III, and IV,
reducing the state space to \(2^3 = 8\) possible states per side, and
\(2 \times 8 \times 8 = 128\) states in total. LNLs I, V, and VII are
excluded, along with their spread parameters, while remaining parameters
are set to their mean values from table~\ref{tbl-midline-params}.

\begin{figure}

\centering{

\pandocbounded{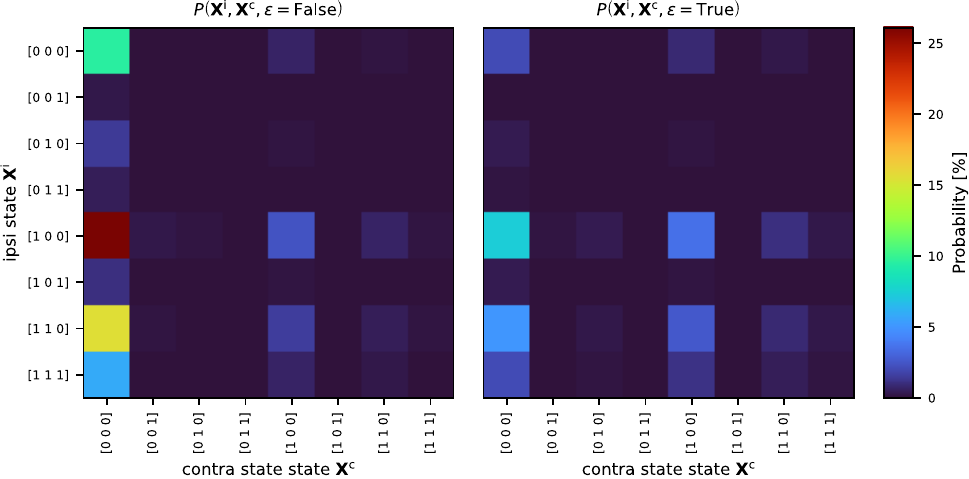}

}

\caption{\label{fig-model-state-dist}The joint distribution
\(P \left( \mathbf{X}^\text{i}, \mathbf{X}^\text{c} \right)\) over ipsi-
and contralateral states and midline extension for late T-category
tumors. The distribution is shown as two separate heatmaps for the two
binary values of the midline extension variable \(\epsilon\). These
matrices are the result of equation~\ref{eq-midline-marginal-algebra},
visualized in figure~\ref{fig-model-algebra-equation}.}

\end{figure}

The left matrix in figure~\ref{fig-model-algebra-equation} shows the
time evolution of the probability distribution over the ipsilateral
involvement states, starting from the healthy state \([0,0,0]\). The two
right matrices show the contralateral state evolution, distinguishing
between the cases of midline extension and no midline extension. At
\(t=0\), the contralateral neck begins in the healthy state \([0,0,0]\)
without midline extension. The central matrix shows the time prior for
late T-category tumors. The matrix multiplication results in the joint
distribution
\(P \left( \mathbf{X}^\text{i}, \mathbf{X}^\text{c} \right)\),
visualized in figure~\ref{fig-model-state-dist}.

This joint distribution is presented as two heatmaps, corresponding to
the two states of midline extension. The most likely state is a
lateralized tumor with ipsilateral level II involvement and no
contralateral involvement, having a probability of approximately 25\%.
The next most probable state is a lateralized tumor with ipsilateral
levels II and III involved, but without contralateral involvement. The
most likely state with contralateral involvement corresponds to tumors
with midline extension, showing involvement of contralateral level II
and ipsilateral levels II and III.

\subsection{Prevalence predictions for contralateral
involvement}\label{prevalence-predictions-for-contralateral-involvement}

The bilateral model was designed to meet the requirements outlined in
section~\ref{sec-requirements}. Here, we evaluate the model's ability to
quantitatively capture the observed patterns of lymph node involvement
in the dataset. Specifically, we compare the model's predictions for
contralateral involvement to the observed data across scenarios that
vary by T-category, midline extension, and ipsilateral involvement.

\subsubsection{Dependence of Contralateral Involvement on T-Category and
Midline
Extension}\label{dependence-of-contralateral-involvement-on-t-category-and-midline-extension}

\begin{figure}

\centering{

\pandocbounded{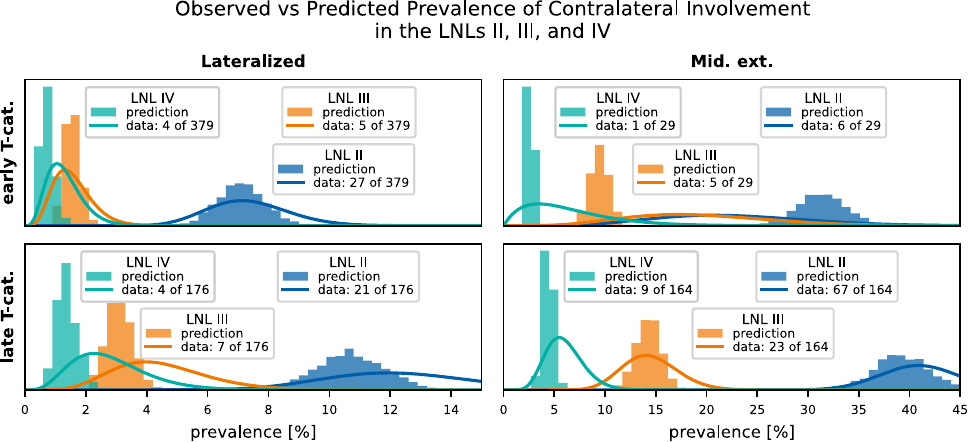}

}

\caption{\label{fig-model-prevalences-overall}Comparison of predicted
(histograms) vs observed (beta posteriors) prevalences, shown for the
contralateral LNLs II (blue), III (orange), and IV (green). The top row
shows scenarios with early T-category tumors, the bottom row for late
T-category tumors. The left column depicts scenarios where the primary
tumor is clearly lateralized, the right column scenarios of tumors
extending over the mid-sagittal plane. This figure illustrates the
model's ability to describe the prevalence of involvement for different
combinations of the risk factors T-category and midline extension.}

\end{figure}

In figure~\ref{fig-model-prevalences-overall}, we compare the prevalence
of contralateral involvement for LNLs II, III, and IV.

Figure~\ref{fig-model-prevalences-overall} demonstrates the model's
ability to account for the risk factors T-category and midline
extension. Consistent with the data, the model predicts that the
prevalence of contralateral LNL II involvement increases from 7.1\% for
early T-category lateralized tumors to 39.2\% for advanced T-category
tumors that cross the midline. Similarly, contralateral LNL III
involvement rises from around 1.5\% for early T-category lateralized
tumors to nearly 14.2\% for advanced T-category midline-extending
tumors.

\subsubsection{Influence of Upstream Involvement on Contralateral
Metastasis}\label{influence-of-upstream-involvement-on-contralateral-metastasis}

\begin{figure}

\centering{

\pandocbounded{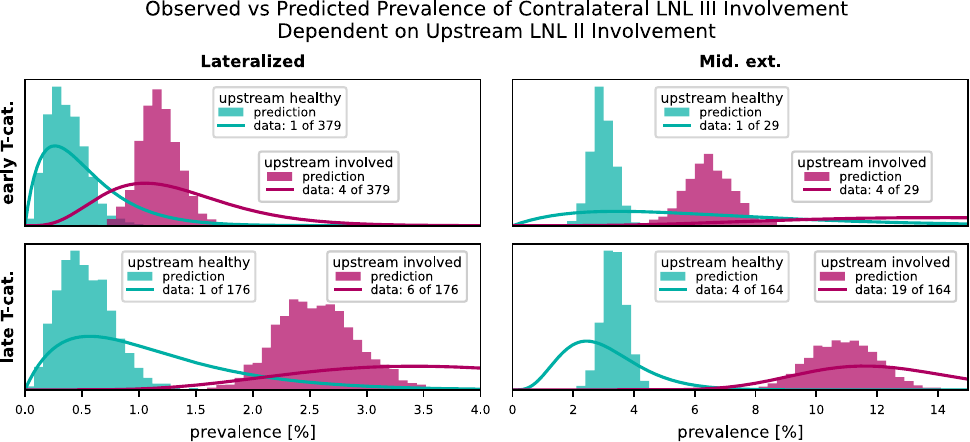}

}

\caption{\label{fig-model-prevalences-upstream}The influence of the
upstream LNL II's involvement on the prevalence of contralateral level
III for the four combinations of tumor lateralization (lateralized or
extending over midline) and T-category (early or advanced). Our model
predictions (histograms) are plotted against the observations in the
data (beta posteriors).}

\end{figure}%

Figure~\ref{fig-model-prevalences-upstream} highlights the influence of
upstream LNL II involvement on contralateral LNL III metastasis. The
contralateral LNL III rarely harbors metastases when its upstream LNL II
is healthy, a correlation well-captured by the model. Out of 164
patients with an advanced tumor crossing the midline and involvement in
the upstream LNL II, 19 had involvement in the contralateral LNL III. In
contrast, out of 379 patients with a clearly lateralized early
T-category tumor and no upstream involvement, only 1 showed
contralateral LNL III involvement. This is well captured by our model.

\subsubsection{Correlation between Ipsi- and Contralateral
Involvement}\label{correlation-between-ipsi--and-contralateral-involvement}

\begin{figure}

\centering{

\pandocbounded{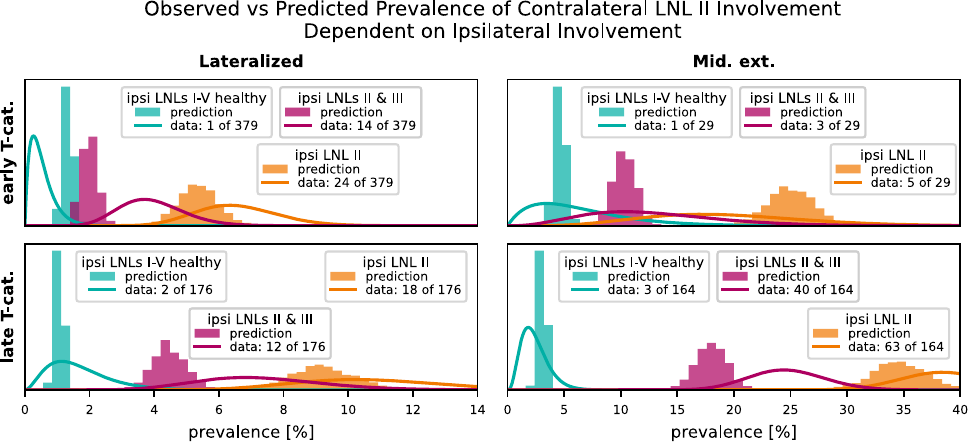}

}

\caption{\label{fig-model-prevalences-with-ipsi}Comparison of the
computed and observed prevalences for scenarios that illustrate the
model's capability of accounting for the correlation between ipsi- and
contralateral involvement. We show three scenarios where we consider the
joint involvement of contralateral LNL II together with different
ipsilateral involvements: 1) the ipsilateral neck shows no involvement
in green (LNLs I to V are healthy, LNL VII is unspecified because data
on it is missing for some patients), 2) where ipsilateral LNL II is
involved in orange (LNLs I, III, IV, and V are healthy), and 3) where
ipsilateral LNLs II and III are involved in red (LNLs I, IV, and V are
healthy). These three scenarios are plotted for all combinations of
T-category (early in top row, advanced in bottom row) and tumor
lateralization (lateralized in left column, extending over mid-sagittal
plane in the right column).}

\end{figure}%

In figure~\ref{fig-model-prevalences-with-ipsi}, the model's ability to
capture the correlation between ipsi- and contralateral involvement is
demonstrated. The marginals of the joint distribution highlight
contralateral LNL II involvement alongside varying ipsilateral LNL
involvement states. Despite having no direct connections between the two
sides, the model successfully predicts these correlations, which arise
purely through the shared diagnosis time.

For example, the model accurately predicts that contralateral LNL II
involvement is rare when the ipsilateral neck is completely healthy
(green histograms). However, if ipsilateral LNL II is involved,
contralateral involvement becomes more likely. Notably, the model
achieves this without specific parameters to quantify ipsi-contralateral
correlations, relying instead on the inherent structure of the
time-dependent dynamics.

\section{Results: Prediction of Risk for Occult
Disease}\label{sec-results-risk}

For clinical applications, the model needs to estimate the risk of
occult metastases in clinically negative LNLs based on a patient's
diagnosis. The diagnosis includes the T-category, tumor lateralization,
and the clinically detected involvement of LNLs based on imaging and
possibly fine needle aspiration (FNA).

We assume imaging detects lymph node involvement with a sensitivity of
81\% and a specificity of 76\% \citep{debondt_detection_2007}, while FNA
has a sensitivity of 80\% and a specificity of 98\%. This implies that
FNA-confirmed involvement is highly reliable, with almost no false
positives.

\begin{figure}

\centering{

\def\svgwidth{0.5\linewidth}
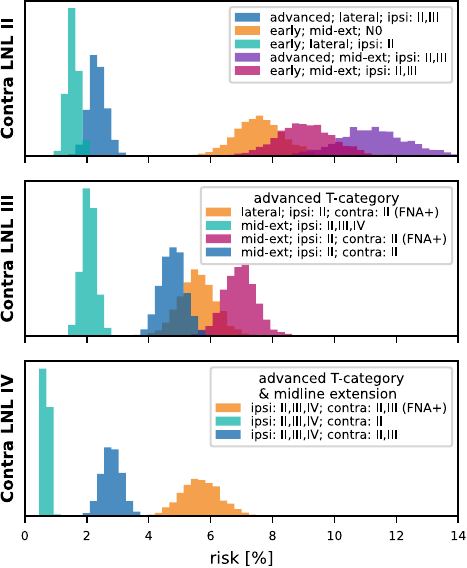

}

\caption{\label{fig-model-risks}Histograms over the predicted risk of
occult involvement in contralateral LNL II (top), III (middle), and IV
(bottom), shown for various combinations of T-category, tumor
lateralization, and clinical LNL diagnoses. All LNLs not explicitly
mentioned in the legend, including the LNL for which the risk of occult
disease was computed, were assumed to be clinically negative
(specificity 76\%, sensitivity 81\%).}

\end{figure}%

\subsection{Contralateral LNL II}\label{contralateral-lnl-ii}

The predicted risk of occult disease in contralateral LNL II is shown in
the left panel of figure~\ref{fig-model-risks}. Tumor lateralization is
the strongest determinant of risk:

\begin{itemize}
\tightlist
\item
  A patient with a lateralized early T-category tumor and ipsilateral
  LNL II involvement has a predicted risk of 1.6\% for occult
  contralateral LNL II disease (green histogram).\\
\item
  For an early T-category tumor that extends over the midline, with
  ipsilateral LNL II involved, the risk increases to 7.6\% (orange
  histogram).
\end{itemize}

Advanced T-category further increases risk but has less impact than
midline extension. For instance:

\begin{itemize}
\tightlist
\item
  An early T-category tumor crossing the midline with ipsilateral
  involvement of LNLs II and III has a 9.1\% risk of contralateral LNL
  II disease (red histogram).\\
\item
  For the same scenario but an advanced T-category tumor, the risk rises
  to 11.3\% (purple histogram).
\end{itemize}

The degree of ipsilateral involvement also influences risk. For a
midline-extending early T-category tumor, and a clinically N0
ipsilateral neck, the predicted risk is 7.6\%. This increases to 9.1\%
when LNLs II and III are involved.

In summary, midline extension is the primary risk factor for
contralateral LNL II involvement, but advanced T-category and extensive
ipsilateral involvement also contribute.

\subsection{Contralateral LNL III}\label{contralateral-lnl-iii}

As shown in the center panel of figure~\ref{fig-model-risks}, the risk
of occult contralateral LNL III involvement rarely exceeds 5\% and
depends strongly on upstream LNL II involvement:

\begin{itemize}
\tightlist
\item
  For an advanced T-category tumor extending over the midline and with
  extensive clinical involvement in the ipsilateral LNLs II, III, and
  IV, but a clinically negative contralateral neck, the risk is only
  2.1\% (green histogram).\\
\item
  If contralateral LNL II is clinically involved, the risk for
  contralateral LNL III rises to 4.8\% (blue histogram).\\
\item
  When contralateral LNL II involvement is confirmed by FNA, the risk
  further increases to 6.9\% (red histogram).
\end{itemize}

Even for lateralized tumors, FNA-confirmed involvement in contralateral
LNL II predicts a 5.6\% risk for LNL III involvement (orange histogram).
This highlights the importance of upstream LNL II involvement in
determining the risk for LNL III.

\subsection{Contralateral LNL IV}\label{contralateral-lnl-iv}

For contralateral LNL IV, the predicted risk is below 1\% in most
scenarios, even for advanced T-category tumors extending over the
midline with extensive ipsilateral and contralateral involvement (green
histogram).

\begin{itemize}
\tightlist
\item
  If contralateral LNL III is also clinically involved, the risk
  increases to 2.8\% (blue histogram).\\
\item
  When contralateral LNL III involvement is confirmed by FNA, the risk
  rises significantly to 5.7\% (orange histogram).
\end{itemize}

This higher risk occurs because FNA confirmation eliminates the
possibility of false-positive diagnoses for contralateral LNL III,
strongly increasing the likelihood of downstream LNL IV involvement.

\subsection{Contralateral LNLs I, V, and
VII}\label{contralateral-lnls-i-v-and-vii}

Contralateral LNLs I, V, and VII show very low predicted risks for
occult disease. Even in advanced T-category tumors extending over the
midline with extensive ipsi- and contralateral involvement, the risk for
LNLs I and VII remains below 3\%.

For contralateral LNL V, the risk is very small unless there is severe
contralateral involvement, including LNL IV, confirmed by FNA. In the
extreme case of advanced T-category tumors extending over the
mid-sagittal plane, with all ipsilateral LNLs, as well as contralateral
LNLs II, III, and IV involved, the risk increases only to 5.5\%.

\section{Discussion}\label{sec-discussion}

\subsection{Summary}\label{summary}

This work introduces a formalism to model ipsi- and contralateral lymph
node involvement in oropharyngeal SCC patients. Building on a previously
developed ipsilateral model
\citep{ludwig_hidden_2021, ludwig_modelling_2023}, we extend it to the
contralateral side while preserving the original model's structure. Our
extension is both intuitive and interpretable, with parameters learned
from a dataset of 833 patients across four institutions. The model uses
clinically diagnosed LNL involvement, tumor T-category, and
lateralization to provide personalized risk predictions for occult
disease in any LNL of interest.

The model is highly interpretable, with each parameter having a clear,
intuitive explanation. Despite its relatively few parameters, it
adequately describes the observed data on ipsilateral and contralateral
nodal involvement. To our knowledge, this represents the most
comprehensive model of lymphatic tumor progression in oropharyngeal SCC,
surpassing prior efforts, which were conceptually different, limited in
scope, or not trained on real patient data
\citep{benson_markov_2006, jung_development_2017}. The underlying
dataset and code are publicly available, supporting reproducibility and
further development.

\subsection{Implications for Contralateral Elective Nodal
Treatment}\label{implications-for-contralateral-elective-nodal-treatment}

Based on a 5\% acceptable risk threshold for occult disease in a LNL,
the model recommends the following approach for contralateral elective
irradiation, assuming the respective LNL is clinically healthy:

\begin{itemize}
\tightlist
\item
  \textbf{Lateralized tumors with no contralateral clinical
  involvement:}\\
  Unilateral radiotherapy is sufficient, regardless of T-category or
  ipsilateral involvement.\\
\item
  \textbf{Tumors extending over the midline with no contralateral
  clinical involvement:}\\
  Elective irradiation is limited to LNL II.\\
\item
  \textbf{Contralateral LNL III:}\\
  Irradiate LNL III if LNL II is involved, regardless of tumor
  lateralization, T-category, or ipsilateral involvement. If
  contralateral LNL II is clinically negative, LNL III is not irradiated
  unless contralateral LNL IV is involved.\\
\item
  \textbf{Contralateral LNL IV:}\\
  Irradiate only when LNL III involvement is confirmed.\\
\item
  \textbf{Contralateral LNL V:}\\
  Elective irradiation is not recommended in almost all patients. Only
  in extreme cases, such as advanced T-category tumors with midline
  extension and confirmed contralateral involvement in LNLs II to IV,
  irradiation of LNL V may be considered.\\
\item
  \textbf{Contralateral LNLs I and VII:}\\
  Elective irradiation is not recommended unless these levels are
  clinically involved.
\end{itemize}

These recommendations align with the model results discussed in
section~\ref{sec-results-risk}, providing a basis for refined treatment
guidelines. However, they should be interpreted in light of the
limitations discussed in section~\ref{sec-discussion-limitations} below.

The model's predictions are already guiding a clinical trial on volume
de-escalation at the University Hospital Zurich
\citep{universityofzurich_personalized_2024}.

\subsection{Limitations and Future
Work}\label{sec-discussion-limitations}

\subsubsection{T-Category Dependence}\label{t-category-dependence}

The model uses a single parameter, \(p_\text{adv.}\), to account for
differences in lymph node involvement patterns between early and
advanced T-category tumors. Advanced T-category tumors are modeled as
evolving over more time steps, while the probability of spread per time
step, governed by the \(b_v\) parameters, remains constant. While this
approach captures the overall differences between early and advanced
T-category tumors well (as seen in
figure~\ref{fig-model-prevalences-overall}), it is not perfect:

\begin{itemize}
\tightlist
\item
  The observed differences between early and advanced T-category tumors
  may sometimes be greater or smaller than the model's predictions. This
  has, for example, been previously noted for ipsilateral LNL I
  involvement \citep{ludwig_modelling_2023}.
\item
  In the context of the bilateral model, the prevalence of midline
  extension is overestimated for early T-category tumors and
  underestimated for advanced T-category tumors, as discussed in
  section~\ref{sec-prevalence-midext}.
\end{itemize}

\subsubsection{Sensitivity and
Specificity}\label{sensitivity-and-specificity}

As noted in section~\ref{sec-training-data} and further detailed in
section~\ref{sec-consensus}, we assumed that the consensus across
diagnostic modalities reflects the true state \(X_v\) of lymph node
involvement. While this assumption is reasonable for pathologically
confirmed diagnoses, it is an approximation for clinically diagnosed
involvement, which cannot detect occult disease by definition.

The model could, in principle, distinguish between pathologically
confirmed and clinically diagnosed involvement by incorporating
different sensitivity and specificity values for each diagnostic
modality during training. However, this was not implemented in the
current work for two reasons:

\begin{enumerate}
\def\labelenumi{\arabic{enumi}.}
\tightlist
\item
  \textbf{Simplified evaluation:} This approximation allowed for direct
  comparison between the observed prevalence of involvement and the
  model's predictions, enabling an evaluation of the model's ability to
  describe the data with few interpretable parameters.
\item
  \textbf{Inconsistent literature values:} Reported sensitivity and
  specificity values for diagnostic modalities show inconsistencies with
  some observed data, complicating their integration into the model.
\end{enumerate}

Future efforts could focus on developing methods to rigorously
differentiate between pathologically confirmed and clinically diagnosed
involvement.

\subsubsection{Tumor Subsites}\label{tumor-subsites}

Oropharyngeal tumors occur in distinct subsites such as the base of the
tongue or the tonsils, which may exhibit slightly different lymphatic
metastatic spread. Incorporating subsite-specific information into the
model could enhance its predictive accuracy. Preliminary investigations
suggest that a mixture model may effectively capture these
subsite-specific spread patterns
\citep{sarrut_proceedings_2024, ludwig_modelling_2023a}.

This approach could also facilitate extending the model to other tumor
locations, such as the oral cavity, hypopharynx, and larynx. Including
these additional tumor sites would broaden the model's applicability to
all HNSCC patients.

\clearpage

\appendix

\section{Acknowledgement}\label{acknowledgement}

We want to thank the research group around Vincent Grégoire from the
Centre Léon Bérard in Lyon (France) and Roland Giger's group from the
Inselspital in Bern (Switzerland) for collecting, curating, and
publishing their data on lymphatic progression patterns. Without their
pathologically assessed data on nodal involvement, we could not have
trained and validated our model with a large and representative cohort.

Further, this work was supported by:

\begin{itemize}
\tightlist
\item
  the Clinical Research Priority Program ``Artificial Intelligence in
  Oncological Imaging'' of the University of Zurich
\item
  the Swiss Cancer Research Foundation under grant number KFS
  5645-08-2022
\end{itemize}

\section{Consensus on most likely involvement}\label{sec-consensus}

The consensus on the most likely involvement state of a LNL was formed
as follows: Suppose the involvement status \(X_v\) of LNL \(v\) was
assessed using different diagnostic modalities
\(\mathcal{O} = \{ \text{MRI}, \text{CT}, \text{pathology}, \ldots \}\),
each characterized by their own pair of sensitivity and specificity
values \(s_N^{o}\) and \(s_P^{o}\), with
\(o \in \mathcal{O}\). These values are tabulated in
table~\ref{tbl-spec-sens}. Then we have \(|\mathcal{O}|\) observations
\(z_v^{o} \in \left[ 0, 1 \right]\), where 0 stands for
``healthy'' and 1 for ``involved''. We can then compute the most likely
true involvement \(X_v\) using the likelihood function
\[
\begin{aligned}
\ell \left( X_v \mid \{ z_v^{o} \}_{o \in \mathcal{O}} \right) = \prod_{o \in \mathcal{O}}
\left( 1 - X_v \right) \cdot &\left[ z_v^{o} \cdot \left( 1 - s_P^{o} \right) + \left( 1 - z_v^{o} \right) \cdot s_P^{o} \right] \\
+ X_v \cdot &\left[ z_v^{o} \cdot s_N^{o} + \left( 1 - z_v^{o} \right) \cdot (1 - s_N^{o}) \right]
\end{aligned}
\]

We now assume the true state \(X_v\) to take on the value 1 if
\(\ell \left( X_v = 1 \mid \ldots \right) > \ell \left( X_v = 0 \mid \ldots \right)\)
and 0 otherwise. For example, if we have \(z_\text{II}^\text{CT} = 0\)
and \(z_\text{II}^\text{MRI} = 1\) we would compute the following
likelihoods:
\[
\begin{aligned}
\ell \left( X_\text{II} = 1 \mid z_\text{II}^\text{CT} = 0, z_\text{II}^\text{MRI} = 1 \right) &= \left( 1 - s_N^\text{CT} \right) \cdot s_N^\text{MRI} = 15.39\% \\
\ell \left( X_\text{II} = 0 \mid z_\text{II}^\text{CT} = 0, z_\text{II}^\text{MRI} = 1 \right) &= s_P^\text{CT} \cdot \left(1 - s_N^\text{MRI}\right) = 14.44\%
\end{aligned}
\]

In this example, we would thus assume the true state to be involved
(\(X_\text{II} = 1\)).

This method of computing a consensus also ensures that the pathology
reports always override any conflicting clinical diagnosis, due to
pathology's high sensitivity and specificity.

\begin{longtable}[]{@{}lrr@{}}
\caption{Specificity and sensitivity values from the literature
\citep{debondt_detection_2007, kyzas_18ffluorodeoxyglucose_2008}.}\label{tbl-spec-sens}\tabularnewline
\toprule\noalign{}
Modality & Specificity & Sensitivity \\
\midrule\noalign{}
\endfirsthead
\toprule\noalign{}
Modality & Specificity & Sensitivity \\
\midrule\noalign{}
\endhead
\bottomrule\noalign{}
\endlastfoot
CT & 76\% & 81\% \\
PET & 86\% & 79\% \\
MRI & 63\% & 81\% \\
FNA & 98\% & 80\% \\
pathology & 100\% & 100\% \\
\end{longtable}

\section{Contralateral Prevalence of
Involvement}\label{contralateral-prevalence-of-involvement}

\begin{longtable}[]{@{}lrlrrrrrrrrr@{}}

\caption{\label{tbl-data-strat}Contralateral involvement depending on
whether the primary tumor extends over the mid-sagittal plane, the
T-category, and how many ipsilateral LNLs were involved.}

\tabularnewline

\tabularnewline
\toprule\noalign{}
T-cat. & ipsi & Mid. ext. & \multicolumn{2}{l}{%
I} & \multicolumn{2}{l}{%
II} & \multicolumn{2}{l}{%
III} & \multicolumn{2}{l}{%
IV} & total \\
& & & n & \% & n & \% & n & \% & n & \% & n \\
\midrule\noalign{}
\endfirsthead
\toprule\noalign{}
T-cat. & ipsi & Mid. ext. & \multicolumn{2}{l}{%
I} & \multicolumn{2}{l}{%
II} & \multicolumn{2}{l}{%
III} & \multicolumn{2}{l}{%
IV} & total \\
& & & n & \% & n & \% & n & \% & n & \% & n \\
\midrule\noalign{}
\endhead
\bottomrule\noalign{}
\endlastfoot
early & 0 & False & 0 & 0.0 & 1 & 1.2 & 0 & 0.0 & 0 & 0.0 & 86 \\
early & 0 & True & 0 & 0.0 & 1 & 10.0 & 1 & 10.0 & 0 & 0.0 & 10 \\
early & 0 & unknown & 0 & 0.0 & 0 & 0.0 & 0 & 0.0 & 0 & 0.0 & 12 \\
early & 1 & False & 1 & 0.5 & 11 & 5.8 & 2 & 1.1 & 1 & 0.5 & 189 \\
early & 1 & True & 1 & 11.1 & 2 & 22.2 & 0 & 0.0 & 0 & 0.0 & 9 \\
early & 1 & unknown & 0 & 0.0 & 3 & 9.7 & 0 & 0.0 & 1 & 3.2 & 31 \\
early & $\geq$ 2 & False & 1 & 1.0 & 15 & 14.4 & 3 & 2.9 & 3 & 2.9 &
104 \\
early & $\geq$ 2 & True & 0 & 0.0 & 3 & 30.0 & 4 & 40.0 & 1 & 10.0 &
10 \\
early & $\geq$ 2 & unknown & 0 & 0.0 & 3 & 13.6 & 0 & 0.0 & 0 & 0.0 & 22 \\
advanced & 0 & False & 0 & 0.0 & 2 & 5.9 & 0 & 0.0 & 0 & 0.0 & 34 \\
advanced & 0 & True & 0 & 0.0 & 3 & 12.5 & 0 & 0.0 & 0 & 0.0 & 24 \\
advanced & 0 & unknown & 0 & 0.0 & 0 & 0.0 & 0 & 0.0 & 0 & 0.0 & 3 \\
advanced & 1 & False & 0 & 0.0 & 3 & 4.6 & 0 & 0.0 & 0 & 0.0 & 66 \\
advanced & 1 & True & 1 & 1.6 & 18 & 29.5 & 5 & 8.2 & 1 & 1.6 &
61 \\
advanced & 1 & unknown & 0 & 0.0 & 0 & 0.0 & 0 & 0.0 & 0 & 0.0 & 9 \\
advanced & $\geq$ 2 & False & 4 & 5.3 & 16 & 21.1 & 7 & 9.2 & 4 & 5.3 &
76 \\
advanced & $\geq$ 2 & True & 3 & 3.8 & 46 & 58.2 & 18 & 22.8 & 8 & 10.1 &
79 \\
advanced & $\geq$ 2 & unknown & 0 & 0.0 & 0 & 0.0 & 0 & 0.0 & 0 & 0.0 & 8 \\

\end{longtable}

In table~\ref{tbl-data-strat}, we show the contralateral involvement of all oropharyngeal SCC patients in our data, stratified by all combinations of T-category, number of ipsilaterally involved LNLs, and whether the primary tumor crossed the mid-sagittal plane.

\section{Prevalence of Midline Extension}\label{sec-prevalence-midext}

\begin{figure}

\centering{

\def\svgwidth{0.5\linewidth}
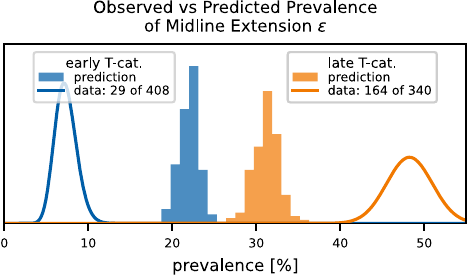

}

\caption{\label{fig-model-prevalences-midext}Comparing the predicted
(histograms) and observed (lines depicting beta posteriors) prevalence
of midline extension for early (blue) and late (orange) T-category.
While the prevalence is predicted correctly when marginalizing over
T-category, the model cannot capture the degree of separation observed
in the data. Since the tumor's midline extension is virtually always
part of the diagnosis and hence \emph{given} when predicting a patient's
risk, we do not consider this discrepancy a major issue.}

\end{figure}%

In figure~\ref{fig-model-prevalences-midext}, we plot the prevalence of
midline extension in the data versus our model's prediction. The model
adjusts the parameter \(p_\epsilon\) to correctly predict the overall
proportion of patients with midline extension, but it cannot match the
large spread between early and advanced T-category seen in the data. To
achieve that, the model would need to increase \(p_\text{adv.}\) and
decrease \(p_\epsilon\). But since the parameter \(p_\text{adv.}\) also
determines the differences in LNL involvement between early and advanced
T-category, the model does not have that freedom.

However, we do not consider this discrepancy a major limitation of the
model: In a practical application of the model we are not interested in
the probability of midline extension, as it is always possible to assess
it with high certainty for the patient at hand. That is also the reason
why we initially modelled the midline extension \emph{not} as a random
variable, but as a global risk factor that would have been turned on or
off from the onset of a patient's disease evolution. This, however, lead
to overly high risks for contralateral involvement in advanced
T-category patients with midline extension, because then the model
assumes an increased spread to the contralateral side from the onset of
the disease. Which is probably not true in a majority of those cases.
Thus, treating it as a random variable that only becomes true during a
patient's disease evolution resulted in a better description of the
data.

Formally, the wrong prediction of midline extension prevalence makes
little difference, since it is always given: Instead of
\(P\left( \mathbf{X}^\text{i}, \mathbf{X}^\text{c}, \epsilon \mid \mathbf{Z}^\text{i}, \mathbf{Z}^\text{c} \right)\),
we typically compute
\(P\left( \mathbf{X}^\text{i}, \mathbf{X}^\text{c} \mid \mathbf{Z}^\text{i}, \mathbf{Z}^\text{c}, \epsilon \right)\),
which does not suffer from the wrong probability of midline extension,
as the distribution over hidden states is renormalized:

\[
P \left( \mathbf{X}^\text{i}, \mathbf{X}^\text{c} \mid \mathbf{Z}^\text{i}, \mathbf{Z}^\text{c}, \epsilon \right) = \frac{P \left( \mathbf{Z}^\text{i}, \mathbf{Z}^\text{c} \mid \mathbf{X}^\text{i}, \mathbf{X}^\text{c}, \epsilon \right) P \left( \mathbf{X}^\text{i}, \mathbf{X}^\text{c}, \epsilon \right)}{P \left( \mathbf{Z}^\text{i}, \mathbf{Z}^\text{c}, \epsilon \right)}
\]

Note that a distribution over \(\epsilon\) appears both in the
enumerator and the denominator, which largely cancel each other, leaving
only the midline extension's effect on the distribution over hidden
states in the prediction.

\bibliography{manuscript.bib}

\end{document}